\documentclass[useAMS,usenatbib]{mn2e}

\usepackage[pdftex]{graphicx}
\usepackage{amsmath}

\usepackage{relsize}

\newcommand{\hi}{%
  \relax
  \ifmmode
    \textrm{\textsc{HI}}
  \else
    \textsc{H{\smaller}I}
  \fi
}

\newcommand{\simgt}{\lower.5ex\hbox{$\; \buildrel > \over \sim \;$}}
\newcommand{\simlt}{\lower.5ex\hbox{$\; \buildrel < \over \sim \;$}}

   \title[Radial Surface Photometry Profiles]{The Way We Measure: Comparison of Methods to Derive Radial Surface Brightness Profiles}
\author[S. P. C. Peters et al.]{S. P. C. Peters$^{1}$,
P. C. van der Kruit$^{1}$\thanks{For more information, please contact P.C. van der Kruit by  email at vdkruit@astro.rug.nl..}, 
R. S. de Jong$^{2}$\\
$^{1}$Kapteyn Astronomical Institute, University of Groningen, P.O.Box 800, 9700AV Groningen, the Netherlands\\
$^{2}$Leibniz~Institut~f\"ur~Astrophysik~Potsdam (AIP), An~der~Sternwarte 16, 14482~Potsdam, Germany.\\
}
\begin{document}
\date{Accepted 2015 month xx. Received 2015 Month xx; in original form 2015 Month xx}
\pagerange{\pageref{firstpage}--\pageref{lastpage}} \pubyear{2015}

\maketitle

\label{firstpage}

  \begin{abstract}
   The breaks and truncations in the luminosity profile of face-on spiral 
galaxies offer valuable insights in their formation history.
    The traditional method of deriving the surface photometry profile for 
face-on galaxies is to use elliptical averaging.
    In this paper, we explore the question whether elliptical averaging is the 
best way to do this.
    We apply two additional surface photometry methods, one new: 
principle axis summation, and one old that has become seldom used: 
equivalent profiles. 
    These are compared to elliptically averaged profiles using a set of 29 
face-on galaxies.
    We find that the equivalent profiles match extremely well with 
elliptically averaged profiles, confirming 
    the validity of using elliptical averaging. 
    The principle axis summation offers a better comparison to edge-on galaxies.
\end{abstract}


\section{Introduction}
The surface photometry of a galaxy is the relationship of the radius $R$, seen 
from the centre of a galaxy, with the surface brightness $\mu(R)$. To first 
order, light is tracing mass in a galaxy. It is therefore an interesting tool 
for the study of galaxy dynamics and evolution. The first studies on the 
subject are by \citet{Patterson1940A} and 
\citet{Vaucouleurs1948A,Vaucouleurs1959A}, who noted that the surface 
brightness of the disc of spiral galaxies followed an exponential decline. 
The exponential nature was studied in more detail by \citet{Freeman1970A}, 
who found that there was a second type of profiles that exhibits a break, 
beyond which the brightness decreases more rapidly.

The lines-of-sight in an edge-on galaxy are typically longer than in a 
face-on galaxy. Thus, more stars are sampled by a single line-of-sight 
through an edge-on than through a face-on galaxy at that same (projected) 
radius.
Because of this, it is easier to detect light at larger radii in edge-on 
galaxies than in face-on galaxies.
This allowed \citet{vdk79} to note that in three edge-on galaxies, the 
radius of the stellar disc did not increase with deeper photograp000hic 
exposures. This work was later expanded to a set of eight edge-on 
galaxies for which the three-dimensional light distribution was 
studied in detail. Each of these galaxies has a truncated disc, 
beyond which the intensity rapidly drops to zero, on
average after $4.2\pm0.6$ radial scale lengths
\citep{vanderKruit1981A,vanderKruit1981B,vanderKruit1982A,vanderKruit1982B}. 
The presence of truncations was confirmed by \citet{2000PDL}, who
found however a ratio of trunction radius to exponential scale length of
only $2.9\pm0.7$.

Truncations in face-on galaxies have, at least in our view, not been 
unambiguously identified. \citet{pt06} used the Sloan 
Digital Sky Survey (SDSS) to study a set of 90 face-on late-type galaxies.
\citet{pt06} identified 14 face-on galaxies with truncations.  This result 
has been disputed 
by \citet{vanderKruit2008A}, who argued that these are in fact breaks similar 
to those found by \citet{Freeman1970A}.
\citet{Erwin2008A} studied 66 barred, early-type galaxies and 
\citet{GEAB11} another sample of 47 early-type non-barred spirals. Many
of these inclined 
systems are classified as having `truncations' (increasingly among later types),
but we remain unconvinced that these are equivalent to those in edge-ons and 
not breaks at higher surface brightness levels.  Combining 
Spitzer and near-IR
observations seems to indicate that the break radii correlate with those of
rings, lenses or spiral arms, and not with a sharp outer decline 
\citep{2014Laineetal}. \citet{2008BakTruPo} found 
from a study of radial colour profiles that breaks in the light profiles often
do not correspond to breaks in the apparent total stellar mass surface density,
in fact leaving no feature whatsoever.  
Recently \citet{2013HHE,2016HHE} have initiated
studiss of a large sample of dwarf galaxies; they find many cases of breaks 
that (unlike spirals) remain in stellar surface density profiles.
Exponential gas disks can have a double 
exponential star formation rate, the break radius being related to the 
instability \citep{2006EH}. \citet{Com12} studied 70 edge-on 
galaxies from the Spitzer Survey of Stellar Structure in Galaxies (S$^4$G) 
and found that many edge-ons have truncations, while often more inward 
breaks could be identified, that occured at similar positions as those 
measured in face-on galaxies by \citet{pt06}. 

The view of breaks and truncations as two separate features was 
put forward by 
\citet{mbt12}. In a study of 34 highly inclined
spiral galaxies, they found that the innermost 
break occurs at $\sim\!8\pm1$ kpc and truncations at $\sim\!14\pm2$ kpc in 
galaxies. 
It should be stated that not all workers agree with this point of view. In 
particular \citet{Erwin2008A}, but also \citet{Erwin2005A} and 
\citet{pt06}, argue that
the breaks really correspond to the truncations in edge-on galaxies. We 
disagree, but will return to this subject more extensively in the next paper
in our studies \citep{Peters2015G}.

Anti-truncated profiles, in which the intensity drops less quickly beyond the 
break than it did before the break, have also been discovered 
\citep{Erwin2005A}. We no further address this issue in this paper, but 
will discuss it in more detail in the next paper \citep{Peters2015G}.

Part of the problem in detecting truncations originates in the different ways 
profiles from edge-ons and face-ons are extracted. In edge-on galaxies, the 
surface photometry is defined as the surface brightness along the major axis 
of the galaxy. This light comes from a variety of radii as the line-of-sight 
crosses through the galaxy. In face-on galaxies, the most common way to derive 
profiles is by performing elliptical averaging, such as that offered by the 
\textsc{IRAF} package \texttt{ellipse} \citep{Jedrzejewski1987,Busko1996}. 
Light in such a profile only comes from structures at a single radius. The 
averaging cancels out any local structure, which might be causing the 
truncations in edge-ons \citep{kf11}.

We believe that these local structures are of importance when looking for 
disc truncations. It is therefore interesting to see what the impact of 
ellipse averaging is on profiles, and to explore alternative ways to derive 
such profiles. 
We use two different methods for deriving surface brightness profiles in 
face-on galaxies that should be less sensitive to local structure and 
deviations from circular symmetry: the Principle Axis Summation and the 
Equivalent Profiles.
In Section \ref{sec:PASmethods}, we will detail the inner workings of 
these methods. We will present our sample of face-on galaxies, based 
on a sub-sample of the work by \citet{pt06}, in Section \ref{sec:PASdata}. 
In Section \ref{sec:PASanalysis}, the data will be analyzed and discussed, 
followed by the conclusions in Section \ref{sec:PASdiscussion}. In order to 
conserve trees, the online Appendix  contains tables 
and figures for individual galaxies.

\section{Surface Photometry Methods}\label{sec:PASmethods}
\subsection{Principle Axis Summation}
The active debate over the nature of truncations in edge-ons versus face-ons sparked our interest in developing a new way of measuring the profiles. While attempts have been made to decompose edge-on galaxies into face-ons, such as \citet{vanderKruit1981A}, \citet{Pohlen2007a}, \citet{Pohlen2004}, \citet{Pohlen2007a} and \cite{Com12}, no real attempt has been made to project face-ons into edge-ons. This enticed us to develop this first method, the Principle Axis Summation (PAS). While not a true projection into an edge-on geometry, the PAS results  resemble the edge-on geometry more closely than those using ellipse-fit profiles. 

\begin{figure}
\centering
  \includegraphics[width=0.48\textwidth]{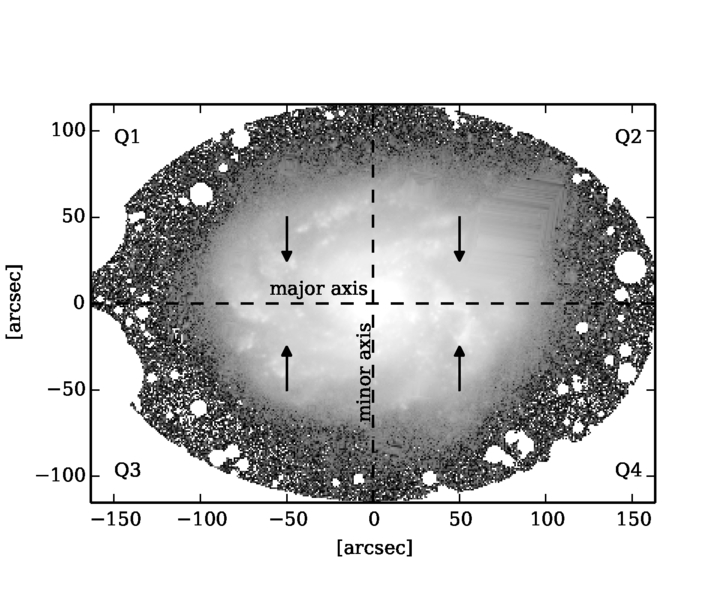}
\caption[Demonstration of the PAS method terminology]{Demonstration of the terminology used in the PAS method. The shown galaxy is NGC\,450. The major and minor axes are shown using the dashed lines. Each quadrant has been labelled. The direction in which the data is summed is shown using the arrows. The outlines of the mask covering background galaxy UGC\,807 are visible in quadrant Q2. This quadrant is therefore ignored in the final PAS analysis. }\label{fig:PASdemo}
\end{figure}

\begin{figure*}
  \includegraphics[width=0.33\textwidth]{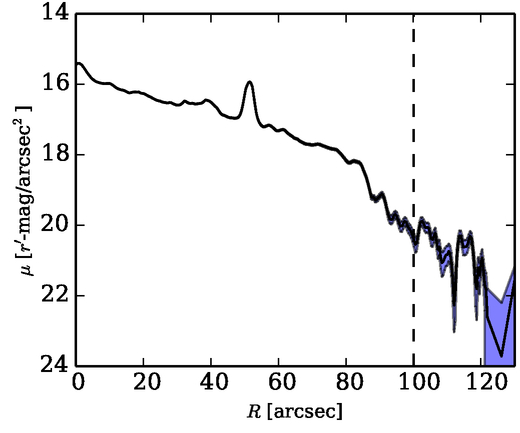}
  \includegraphics[width=0.33\textwidth]{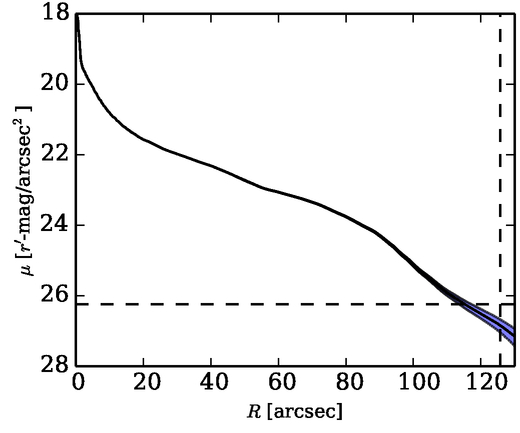}
  \includegraphics[width=0.33\textwidth]{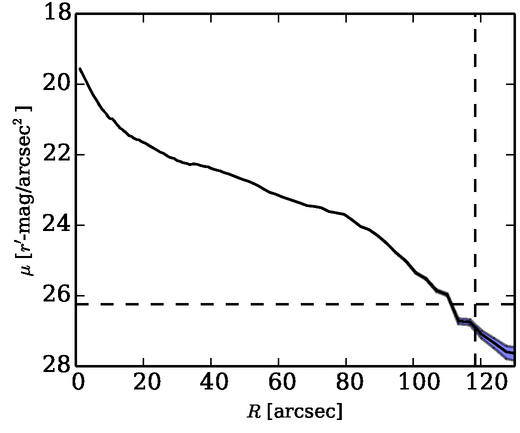}
\caption[Sensitivities of the methods to background estimation issues]{Demonstration of the sensitivity to background under- and overestimation. 
    Here we have used an offset of 0.1 ADU.
    The horizontal, dashed line shows the level of the pixel-to-pixel noise.
    The vertical, dashed line shows the point where the profile deviate by more than 0.2 magnitudes, which represents the point up to which we trust the profile. From left to right the panels correspond to the PAS, EP and ellipse-fit methods.
    The profiles are based on NGC\,450 (see Figure \ref{fig:NGC450}).}
\label{figure:demo}
\end{figure*}

The PAS method partitions the face-on galaxy into four quadrants, centred on 
its major and minor axis. Each quadrant is summed onto the major axis, leaving 
four quadrant-profiles $Q_1(R), Q_2(R), Q_3(R)$ and $Q_4(R)$ (see Figure 
\ref{fig:PASdemo}). These are multiplied by two, to represent the full 
line-of-sight along the major axis and to  represent the line-of-sight 
integration in an edge-on galaxy better. The main profile $P(R)$ is taken 
as the median of these four. The scatter between the four quadrants is 
a good measure of any asymmetry in the galaxy. In cases where one or more 
quadrants suffer from severe contamination by foreground or background 
objects, that quadrant can be ignored and only the clean quadrants will 
be used for the median.
A clear example of this is in {NGC\,450}, where background galaxy {UGC\,807} 
is covering a significant part of a quadrant (Quadrant Q2 in Figure 
\ref{fig:PASdemo}).
We run a dynamic binning algorithm along the main profile, to ensure 
that each point has at least a signal-to-noise ratio of two.
We use an elliptical blanking mask around each galaxy, shaped and 
oriented according to the 25th magnitude ellipse of the galaxy and 
blanked beyond a trust radius $R_\mathrm{t}$, to minimize the 
contribution of sky noise.
The trust radius $R_\mathrm{t}$ is determined by eye on a 
heavily smoothed image, such that the galaxy is fully included in the mask.

The noise in each quadrants profile is a combination of 
the intrinsic pixel-to-pixel noise, any large-scale fluctuations 
and blanked regions.
It thus varies with radius as the amount of pixels in the summation changes.
The main profile depends on the combination of four of these varying 
quadrants and can thus vary drastically.
To have a good representation of the noise levels we calculate the 
noise in the profiles using the sky, 
as taken from the ellipse between one and two times the trust radius 
$R_\mathrm{t}$. 
All pixels between these two radii are selected and merged row by row 
into a single long row of pixels.
For each quadrant, we smooth a copy of this row of sky pixels with a 
`tophat' kernel with the length of the amount of pixels used, effectively 
recreating the pixel summation.
We randomly select a value out of each of these four smoothed sets 
and take the median. 
This is repeated 10000 times and the noise is then calculated as the 
standard deviation of this sample.

There are two major differences between this projection and a true edge-on. 
First, in real edge-ons we would be able to observe the effect of 
variations with height.
As we are seeing the galaxy from above, PAS cannot show this effect.
Compared to ellipse averaging, due to the summation the surface 
brightness in PAS will also be brighter.
The summation effectively has the unit magnitude per arcsec, 
making it distance dependent.
For a true comparison with edge-ons, one would therefore 
need to apply the PAS method to those as well.
The overall shape of the profile should however be equal.
A second major difference is that dust absorption is less of an issue here.
In a true edge-on, this could have a significant impact 
on the scale length of the profile.
This is an interesting feature, as a statistical comparison 
in a large sample of edge-ons and face-ons could be used to 
analyze the dust content of galaxies.

The PAS profiles are more susceptible to sky determination 
issues than ellipse-fit profiles, as any remaining background 
offset will be multiplied by the amount of pixels along the 
minor axis instead of being averaged.
In this paper, we will use the uncertainty in the background-offset 
estimation (see Section \ref{sec:uncertainty}) to over- and 
under-subtract the profile. 
We place our confidence limit at the spot where these three 
profiles start to deviate by more than 0.2 magnitudes.
In Figure \ref{figure:demo} (left), the sensitivity to the 
background is demonstrated, by over- and under-subtracting the data by 0.1 ADU.

As noted before, the PAS profiles are effectively in units of 
magnitude per arcsec. 
Because of this, direct comparison with the other two types of 
profiles is hard. 
Still, we have chosen to  display all these profiles together 
in one graph, by applying an offset to the  PAS profiles, such 
that at $R=0$ the brightest EP profile begins at the same magnitude 
as the faintest PAS profile.
Direct comparisons of the brightness of the PAS profile with 
the EP and ellipse-fit profiles should not be made.
This strategy does however allow for the check if a feature 
occurs at a particular radius $R$ in all three types of profiles.

\subsection{The Equivalent Profiles}\label{chapter:EP}
The Equivalent Profiles (EP) are a radical twist on the usual methods. 
Instead of using the radius $R$ to find the surface brightness $\mu(R)$ in a 
face-on galaxy, the method turns things around.
For each observed surface brightness $\mu$ in the image, there will always be 
some number of pixels $N(\mu)$ that have that or a brighter value.
Since each pixel covers a small surface $dA$, a total equivalent surface 
$A(\mu)$ can be formed.
In SDSS, the area of each pixel covers $0.396\times0.396$ arcsec$^2$.
Assume that the surface brightness in a galaxy is always brightest in the 
centre and decreases with radius\footnote{With the exception of small-scale 
features, this holds for all three types of profiles, the only differences 
between them is the rate at which the brightness decreases.}.
The surface will then form an ellipse, or circle in the case of a perfect 
face-on, centred on the galaxy.
The radius of this equivalent surface is called the equivalent radius $R(\mu)$.
Mathematically we can describe this as 
\begin{equation}
 R(\mu) = \sqrt{\frac{ N(\mu) dA}{\pi \cos i}}\,\,,
\end{equation}
where $i$ is the inclination of the galaxy.

As an example, suppose for a perfectly face-on galaxy that the brightest 
pixel in the observation has a value of $\mu=18$ $r$'-mag/arcsec$^2$.
Since this value is only reached in one pixel, the equivalent area  
$A(\mu)$ is only $0.396^2$\,arcsec$^2$, and the equivalent radius 
$R(\mu)$ is thus $0.35$ arcsec.
At $\mu=$20 $r$'-mag/arcsec$^2$ there could be 10.000 pixels at 
that or a brighter value.
In that case the equivalent area  $A(\mu)$ goes up to 
$0.396^2 \times 10.000 = 1568$\,arcsec$^2$, and the equivalent 
radius $R(\mu)$ is thus 22.3''.
By repeating this process for every value of $\mu$ in the 
observation, we can thus build up the associated set of equivalent radii.

\begin{figure}
 \centering
 \includegraphics[width=0.48\textwidth]{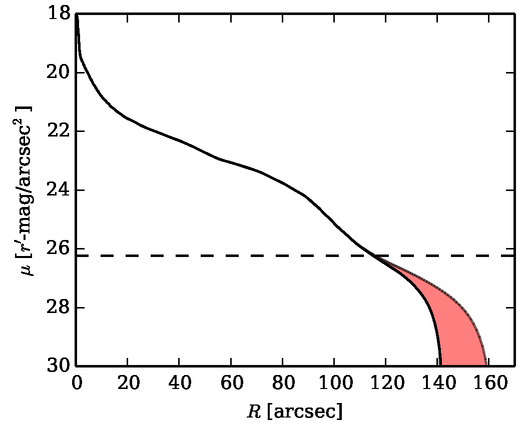}
 \caption[Effective of elliptical mask size of EP results]{Effect of choice of radius for the elliptical mask on the Equivalent Profiles. The radius of the mask has been increased by $25\%$. The shaded region shows the increase in profile compared to the original, black profile.
    The horizontal, dashed line shows the level of the pixel-to-pixel noise.}
 \label{figure:effectofradiuschoice}
 \end{figure}

Tests show that this method is particularly sensitive to background noise.
Any positive component of the noise distribution will add 
to surface $A(\mu)$ and thus increase radius $R(\mu)$.
This creates a drastic increase in the equivalent radius at 
the faintest surface brightness levels (see for example Figure 
\ref{figure:effectofradiuschoice}).
The other methods suffer much less from this, as the positive 
noise values are averaged out against the negative noise values.
Two techniques are used to deal with this. 
Firstly, similar as in the PAS, we use an elliptical blanking 
mask around the galaxy. 
It is centred on the galaxy and has sufficient radius not to  
blank the galaxy itself, but leaves as little background as possible.
This blocks out all signals for which we are sure that they are 
unrelated to the galaxy.
Secondly, we use non-linear anisotropic filtering, an algorithm 
normally used in magnetic resonance imaging \citep{Jones2003A}.
This helps smooth low S/N regions, while conserving the flux and 
important structure in the image.

Equivalent Profiles are an old method, going back more than 60 years. 
The oldest reference traces back to \citet{Vaucouleurs1948A}, 
wherein he derives his famous $R^{1/4}$ profile.
In the decades beyond, they were used quite often, as for example 
in the photometric survey by \citet{vdk79}.
The newer elliptically averaged profiles suffer less from noise, and 
are able to vary the position angle and inclination as function of 
radius  \citep{Jedrzejewski1987}, things that the Equivalent Profiles cannot. 
This is likely why the Equivalent Profiles have fallen from grace.

Similar to the PAS, the confidence limit of the profile is again 
calculated by over- and under-subtracting the data by two times the 
uncertainty and establishing where the profiles start to deviate by 
more than 0.2 magnitudes.
We demonstrate this contamination by background noise in Figure 
\ref{figure:demo} (middle). 
Comparing it to the profiles from the elliptical averaging (reproduced 
here from \citet{pt06}); we see that the Equivalent Profiles start to 
suffer at a brighter magnitude levels.
In practice, this level is slightly higher than the background pixel-to-pixel 
noise $\sigma$. 
The choice of the radius of the mask is also not trivial, as demonstrated in 
Figure \ref{figure:effectofradiuschoice}.
The larger the radius, the more background is sampled, and the more noise is 
picked up.
We have opted to use the same ellipse, with trust radius $R_\textrm{t}$ as used 
for the PAS.

\section{Data}\label{sec:PASdata}
\subsection{Sample}
We use the full sample defined by \citet{pt06}.
They used the following criteria to define their sample:
\vspace{-.2cm}
\begin{itemize}
 \item A Hubble type $T$ parameter between $2.99 < T < 8.49$. This corresponds 
to an intermediate to late-type galaxy sample with Sb to Sdm galaxies.
 \item The axis ratio is chosen such that the inclination is $i <~ 61^\circ$, 
as to avoid the influence of dust and as a convenient way to classify 
morphological properties of the galaxy that would have been more 
obscured at higher inclinations.
 \item The recession velocity is $v_\textrm{vir} < 3250$ km/s and 
the total B-band brightness M$_\textrm{B,abs}< -18.5$ B-mag, as to 
get a complete sample of galaxies within the $~46$\,Mpc survey distance. 
 \item Galactic latitude $\|b_{II}\| > 20^\circ$ as to avoid dust obscuration.
\end{itemize}

\vspace{-0.2cm}
Using DR2 of SDSS \citep{SDSS2}, this led them to a sample of 98 
face-on galaxies for which observations were available, out of a 
full sample of 655 galaxies. The final sample is listed in Table 
\ref{tbl:fundamental}.

\begin{table*}
\begin{tabular}{ llllllll }
\textbf{Galaxy} & \textbf{$M_\textrm{B,abs}$ } & \textbf{Type } & \textbf{t} & \textbf{$v_\textrm{rot} [km/s]$ } & \textbf{$i$ [$^\circ$]} & \textbf{PA [$^\circ$]}  & \textbf{D [Mpc]} \\ \hline
  IC1067                   & -18.65 &  Sb     &    3.0  & 148.74 & 42.3 & 151.1 & 28.3 \\ 
  IC1125                   & -20.03 &  SBcd   &    7.3  & 103.77 & 55.9 & 305.8 & 35 \\ 
  IC1158                   & -19.52 &  SABc   &    5.1  & 120 & 55.9 & 136.7 & 29.7 \\ 
  NGC0450                  & -19.72 &  SABc   &    5.8  & 102.94 & 50.2 & 188.8 & 19 \\ 
  NGC0701                  & -19.84 &  SBc    &    5.0  & 120.96 & 59.3 & 45.7 & 19.5 \\ 
  NGC0853                  & -16.23 &  Sm     &    8.6  & 60.29 & 50.2 & 16.3 & 21 \\ 
  NGC0941                  & -19.13 &  SABc   &    5.3  & 88.93 & 45.6 & 101.4 & 22 \\ 
  NGC1042                  & -20.27 &  SABc   &    6.0  & 46.1 & 36.9 & 74.7 & 4.21 \\ 
  NGC1068                  & -21.5 &  Sb     &     3.0  & 282.54 & 24.5 & 170.2 & 10.1 \\ 
  NGC1084                  & -20.63 &  Sc     &    4.9  & 194.52 & 52.4 & 52.8 & 16.6 \\ 
  NGC1087                  & -20.65 &  SABc   &    5.2  & 120.27 & 52.4 & 268.4 & 19 \\ 
  NGC1299                  & -19.35 &  SBb    &    3.0  & 120.91 & 56.6 & 42.0 & 32 \\ 
  NGC2541                  & -18.66 &  SABc   &    6.0  & 97.22 & 61.3 & 107.6 & 14.8 \\ 
  NGC2543                  & -20.5 &  Sb     &     3.0  & 148.11 & 60.0 & 51.1 & 26.3 \\ 
  NGC2684                  & -19.88 &  Sc     &    4.6  & 101.03 & 34.9 & 35.1 & 44.9 \\ 
  NGC2701                  & -20.45 &  SABc   &    5.2  & 143.88 & 47.2 & 63.3 & 30.7 \\ 
  NGC2776                  & -21.54 &  SABc   &    5.2  & 99.06 & 18.2 & 6.0 & 38.7 \\ 
  NGC2967                  & -20.37 &  Sc     &    5.2  & 165.95 & 21.6 & 250.7 & 30.9 \\ 
  NGC3055                  & -20.12 &  SABc   &    5.3  & 142.65 & 54.5 & 27.0 & 28 \\ 
  NGC3246                  & -19.3 &  Sd     &     7.9  & 109.85 & 58.7 & 354.4 & 35.5 \\ 
  NGC3259                  & -19.62 &  SABb   &    3.7  & 120.54 & 55.2 & 71.5 & 35.9 \\ 
  NGC3310                  & -20.11 &  SABb   &    4.0  & 288.38 & 18.2 & 70.7 & 17.5 \\ 
  NGC3359                  & -20.57 &  Sc     &    5.2  & 148.06 & 58.7 & 101.9 & 22.6 \\ 
  NGC3423                  & -19.6 &  Sc     &    6.0  & 127.12 & 35.9 & 56.3 & 11.7 \\ 
  NGC3488                  & -19.9 &  SBc    &    5.2  & 122.69 & 48.7 & 92.7 & 46.3 \\ 
  NGC3583                  & -20.58 &  SBb    &    3.1  & 182.1 & 47.2 & 326.4 & 31.6 \\ 
  NGC3589                  & -18.63 &  SABc   &    7.0  & 77.82 & 60.0 & 36.3 & 34.1 \\ 
  NGC3631                  & -21.02 &  Sc     &    5.2  & 78.36 & 32.9 & 339.9 & 21.6 \\ 
  NGC3642                  & -20.57 &  Sbc    &    4.0  & 48.71 & 18.2 & 7.4 & 27.5 \\ 
  NGC3756                  & -20.2 &  SABb   &    4.0  & 145.95 & 60.0 & 91.0 & 15.7 \\ 
  NGC3888                  & -20.47 &  SABc   &    5.3  & 203.06 & 42.3 & 335.8 & 41.5 \\ 
  NGC3893                  & -21 &  SABc   &    5.2  & 147.68 & 53.8 & 282.4 & 15.7 \\ 
  NGC3982                  & -19.91 &  SABb   &    3.2  & 191.83 & 27.1 & 92.2 & 24.6 \\ 
  NGC3992                  & -21.31 &  Sbc    &    4.0  & 295.12 & 54.5 & 19.8 & 22.9 \\ 
  NGC4030                  & -20.84 &  Sbc    &    4.0  & 201.32 & 36.9 & 59.6 & 25 \\ 
  NGC4041                  & -20.19 &  Sbc    &    4.0  & 263.1 & 18.2 & 12.4 & 22.7 \\ 
  NGC4102                  & -19.4 &  SABb   &    3.1  & 158.14 & 55.2 & 50.6 & 16 \\ 
  NGC4108                  & -20.25 &  Sc     &    5.2  & 223.28 & 39.6 & 323.1 & 41.6 \\ 
  NGC4108B                 & -18.77 &  SABc   &    7.0  & 195.8 & 38.7 & 349.7 & 43.8 \\ 
  NGC4123                  & -19.91 &  Sc     &    5.0  & 128.5 & 47.9 & 324.0 & 14.9 \\ 
  NGC4210                  & -19.99 &  Sb     &    3.0  & 162.96 & 40.5 & 351.9 & 44.8 \\ 
  NGC4273                  & -20.6 &  Sc     &    5.2  & 328.91 & 52.4 & 263.3 & 28.5 \\ 
  NGC4480                  & -20.3 &  SABc   &    5.1  & 169.24 & 60.0 & 92.6 & 36.7 \\ 
  NGC4517A                 & -19.8 &  Sd     &    7.8  & 71.35 & 55.2 & 241.4 & 23.6 \\ 
  NGC4545                  & -20.3 &  Sc     &    5.6  & 129.19 & 54.5 & 264.7 & 38.2 \\ 
  NGC4653                  & -20.33 &  SABc   &    6.0  & 211.75 & 33.9 & 101.5 & 39.1 \\ 
  NGC4668                  & -18.92 &  SBcd   &    7.4  & 62.33 & 58.7 & 265.5 & 17.2 \\ 
  NGC4904                  & -19.12 &  Sc     &    5.8  & 105.15 & 44.8 & 243.4 & 20.5 \\ 
  NGC5147                  & -19.09 &  SBd    &    7.9  & 154.83 & 37.8 & 150.5 & 21.6 \\ 
  NGC5300                  & -18.7 &  SABc   &    5.2  & 120.42 & 47.9 & 119.6 & 19.9 \\ 
  NGC5334                  & -19.12 &  Sc     &    5.2  & 132.75 & 39.6 & 76.0 & 24.7 \\ 
  NGC5376                  & -20.03 &  SABa   &    2.3  & 204.71 & 52.4 & 208.9 & 55.5 \\ 
  NGC5430                  & -20.76 &  SBb    &    3.1  & 186.86 & 49.5 & 87.4 & 37.9 \\ 
  NGC5480                  & -19.94 &  Sc     &    5.0  & 150.36 & 31.8 & 231.8 & 22.4 \\ 
  NGC5584                  & -19.69 &  SABc   &    6.0  & 124.86 & 42.3 & 292.1 & 19.7 \\ 
  NGC5624                  & -18.75 &  Sbc    &    3.8  & 66.52 & 48.7 & 71.8 & 35 \\ 
  NGC5660                  & -20.66 &  SABc   &    5.2  & 138.52 & 18.2 & 129.6 & 37.2 \\ 
  NGC5667                  & -19.93 &  SBc    &    6.0  & 100.4 & 58.0 & 100.1 & 34.8 \\ 
  NGC5668                  & -20.01 &  Scd    &    6.9  & 72.52 & 31.8 & 134.0 & 26.9 \\ 
  NGC5693                  & -19.08 &  Scd    &    6.9  & 44.83 & 18.2 & 139.6 & 40.1 \\ 
  NGC5713                  & -21.16 &  SABb   &    4.0  & 107.91 & 29.5 & 81.6 & 18.3 \\ 
  NGC5768                  & -19.43 &  Sc     &    5.3  & 123.62 & 27.1 & 337.4 & 33.1 \\ 
\hline
\end{tabular}
\caption{Fundamental properties for the full sample.}
\end{table*}
\label{tbl:fundamental}

\addtocounter{table}{-1}

\begin{table*}
\begin{tabular}{ llllllll }
\textbf{Galaxy} & \textbf{$M_\textrm{B,abs}$ } & \textbf{Type } & \textbf{t } & \textbf{$v_\textrm{rot} [km/s]$}  & \textbf{$i$ [$^\circ$]} & \textbf{PA [$^\circ$]} & \textbf{D [Mpc]}\\ \hline
  NGC5774                  & -19.37 &  SABc   &    6,9  & 83.64 & 38.7 & 135.1 & 26.8 \\ 
  NGC5806                  & -19.92 &  Sb     &    3,2  & 190.93 & 57.3 & 104.7 & 25.2 \\ 
  NGC5850                  & -21.5 &  Sb     &    3,1  & 117.44 & 38.7 & 103.4 & 28.5 \\ 
  NGC5937                  & -21.17 &  SABb   &    3,2  & 180.29 & 55.9 & 70.9 & 46.6 \\ 
  NGC6070                  & -21.14 &  Sc     &    6,0  & 204.85 & 64.5 & 210.8 & 27.8 \\ 
  NGC6155                  & -20.02 &  Sc     &    5,2  & 109.64 & 43.9 & 133.3 & 41.7 \\ 
  NGC7437                  & -18.75 &  SABc   &    6,7  & 151.78 & 25.8 & 70.9 & 29.2 \\ 
  NGC7606                  & -21.2 &  Sb     &    3,0  & 445.74 & 67.7 & 125.7 & 31 \\ 
  PGC006667                & -18.28 &  Scd    &    6,6  & 136.08 & 35.9 & 328.0 & 24.6 \\ 
  UGC02081                 & -18.39 &  SABc   &    5,8  & 95.56 & 54.5 & 108.8 & 42.5 \\ 
  UGC04393                 & -19.23 &  Sbc    &    3,5  & 62.21 & 56.6 & 29.0 & 15.2 \\ 
  UGC06309                 & -19.74 &  SBc    &    4,5  & 134.19 & 46.4 & 326.5 & 47 \\ 
  UGC06518                 & -19.06 &  Sbc    &    3,8  & 87.56 & 52.4 & 68.7 & 46.3 \\ 
  UGC06903                 & -18.35 &  Sc     &    5,9  & 163.69 & 28.4 & 131.9 & 30.5 \\ 
  UGC07700                 & -18.87 &  Sd     &    7,9  & 84.98 & 40.5 & 200.7 & 48.3 \\ 
  UGC08041                 & -18.48 &  SBcd   &    6,9  & 103.6 & 58.7 & 290.0 & 17.2 \\ 
  UGC08084                 & -18.84 &  SBd    &    8,0  & 88.77 & 34.9 & 210.5 & 41.1 \\ 
  UGC08237                 & -19.82 &  SBb    &    3,0  &  & 38.7 & 138.5 & 47 \\ 
  UGC08658                 & -19.93 &  Sc     &    5,0  & 124.79 & 52.4 & 160.3 & 37.2 \\ 
  UGC09741                 & -18.84 &  Sbc    &    4,0  &  & 27.1 & 102.3 & 42.9 \\ 
  UGC09837                 & -19.48 &  SABc   &    5,3  & 179.15 & 18.2 & 117.6 & 41.2 \\ 
  UGC10721                 & -19.72 &  Sc     &    5,8  & 143.07 & 47.9 & 159.8 & 45.8 \\ 
  UGC12709                 & -19.05 &  SABm   &    8,7  & 70.61 & 52.4 & 13.0 & 35.1 \\ 
\hline
\end{tabular}
\caption{Fundamental properties for the full sample, continued.}
\end{table*}

\begin{table*}
\centering
\begin{tabular}{lcclcccc}
\textbf{Galaxy} & \multicolumn{2}{c}{\textbf{Quality}} & \textbf{Type} & \multicolumn{4}{c}{\textbf{Fit radii}} \\ 
 & \textbf{Image} & \textbf{Profile} &  & \textbf{$r_1$} & \textbf{$r_2$} & \textbf{$r_3$} & \textbf{$r_4$} \\ \hline
  IC1067                   & G & M & I & 30 & 80 &   &   \\ 
  IC1125                   & G & G & II & 30 & 40 & 45 & 60 \\ 
  IC1158                   & G & G & II & 40 & 60 & 70 & 100 \\ 
  NGC0450                  & G & G & II & 40 & 75 & 84 & 100 \\ 
  NGC0701                  & G & B & I in ellipse, II in PAS & 20 & 60 & 60 & 80 \\ 
  NGC0853                  & G & B & III in EP en Ell, I in PAS & 25 & 50 & 50 & 70 \\ 
  NGC0941                  & G & G & II & 20 & 65 & 70 & 110 \\ 
  NGC1299                  & G & M & II in PAS, III in EP and ellipse & 0 & 25 & 30 & 70 \\ 
  NGC2701                  & G & G & II and III & 20 & 40 & 50 & 70 \\ 
  NGC2776                  & G & B & II in PAS, III in EP and ellipse & 20 & 80 & 80 & 140 \\ 
  NGC2967                  & G & G & III & 20 & 50 & 90 & 140 \\ 
  NGC3055                  & G & G & II & 20 & 55 & 60 & 80 \\ 
  NGC3259                  & G & G & III & 20 & 40 & 50 & 100 \\ 
  NGC3423                  & G & G & II & 30 & 85 & 100 & 140 \\ 
  NGC3488                  & G & M & II en 3 & 20 & 30 & 40 & 60 \\ 
  NGC3589                  & G & G & II & 10 & 28 & 35 & 50 \\ 
  NGC3631                  & G & M & II (EP fails due to variations) & 55 & 100 & 120 & 170 \\ 
  NGC3642                  & G & M & III & 15 & 75 & 80 & 150 \\ 
  NGC3888                  & G & M & II & 20 & 45 & 45 & 70 \\ 
  NGC3982                  & G & G & III & 20 & 40 & 50 & 80 \\ 
  NGC4041                  & G & G & III & 40 & 70 & 85 & 150 \\ 
  NGC4102                  & G & G & II & 20 & 60 & 60 & 90 \\ 
  NGC4108                  & G & B & II & 10 & 40 & 50 & 60 \\ 
  NGC4108B                 & G & B & I in ellipse, II in PAS & 0 & 40 & 40 & 55 \\ 
  NGC4273                  & G & B & II & 40 & 60 & 80 & 120 \\ 
  NGC4545                  & G & G & II & 20 & 60 & 60 & 70 \\ 
  NGC4653                  & G & G & II & 20 & 80 & 105 & 125 \\ 
  NGC4668                  & G & M & II en 3 & 20 & 33 & 40 & 50 \\ 
  NGC4904                  & G & G & II & 15 & 35 & 50 & 80 \\ 
  NGC5147                  & G & G & II & 7 & 35 & 40 & 70 \\ 
  NGC5300                  & G & G & II & 13 & 75 & 100 & 140 \\ 
  NGC5334                  & G & G & II & 40 & 80 & 100 & 140 \\ 
  NGC5376                  & G & G & II & 10 & 30 & 40 & 70 \\ 
  NGC5430                  & G & G & II & 30 & 48 & 60 & 90 \\ 
  NGC5480                  & G & G & III & 20 & 40 & 80 & 120 \\ 
  NGC5624                  & G & B & III in EP en Ell, I in PAS & 20 & 40 & 40 & 80 \\ 
  NGC5660                  & G & M & II & 20 & 60 & 70 & 80 \\ 
  NGC5667                  & G & B & II (EP fails due to variations) & 10 & 35 & 45 & 60 \\ 
  NGC5668                  & G & G & II & 50 & 80 & 85 & 100 \\ 
  NGC5693                  & G & G & II & 10 & 30 & 40 & 70 \\ 
  NGC5713                  & G & B & I in ellipse, I in PAS & 50 & 85 & 100 & 130 \\ 
  NGC5774                  & G & G & II & 20 & 80 & 85 & 130 \\ 
  NGC5806                  & G & M & III & 50 & 100 & 150 & 200 \\ 
  NGC6155                  & G & M & II & 0 & 30 & 40 & 60 \\ 
  NGC7437                  & G & G & II & 20 & 40 & 50 & 80 \\ 
  PGC006667                & G & G & II & 30 & 70 & 80 & 110 \\ 
  UGC02081                 & G & G & II & 0 & 55 & 60 & 90 \\ 
  UGC04393                 & G & B & II & 40 & 60 & 60 & 70 \\ 
  UGC06309                 & G & B & II & 20 & 40 & 40 & 60 \\ 
  UGC06518                 & G & G & II & 10 & 25 & 25 & 38 \\ 
  UGC06903                 & G & G & II & 20 & 50 & 60 & 95 \\ 
  UGC07700                 & G & G & II & 20 & 39 & 49 & 80 \\ 
  UGC08084                 & G & G & II & 11 & 40 & 45 & 60 \\ 
  UGC08658                 & G & G & II & 20 & 49 & 55 & 90 \\ 
  UGC09741                 & G & G & III & 10 & 20 & 25 & 40 \\ 
  UGC09837                 & G & G & II & 20 & 45 & 53 & 67 \\ 
  UGC12709                 & G & G & II & 31 & 75 & 75 & 100 \\ 
\hline\
\end{tabular}
\caption{\label{tbl:classifications}Quality and fit radii for the approved image sample. Profile quality is split into types bad, moderate and good. Radii in arcsec.}
\end{table*}

\begin{table*}
\centering
{\small
\begin{tabular}{ l|ll|llllll }
\textbf{} & \multicolumn{2}{c|}{\textbf{Ellipse}} & \multicolumn{6}{c}{\textbf{Equivalent Profiles}} \\ 
\textbf{Galaxy} & \textbf{$h_0$ r'} & \textbf{$h_f$ r'}  &\textbf{$h_0$ r'} & \textbf{$h_f$ r'} &\textbf{$h_0$ g'} & \textbf{$h_f$ g'}  & \textbf{$h_0$ i'} & \textbf{$h_f$ i'} \\ \hline
  IC1125                  &$2.64\pm0.05$&$1.63\pm0.06$&$2.48\pm<0.01$&$1.66\pm< 0.01$&$2.72\pm< 0.01$&$1.69\pm< 0.01$&$2.38\pm< 0.01$&$1.95\pm< 0.01$\\
  IC1158                  &$3.53\pm0.10$&$1.49\pm0.03$&$3.23\pm< 0.01$&$1.51\pm< 0.01$&$3.22\pm< 0.01$&$1.66\pm< 0.01$&$3.34\pm< 0.01$&$1.69\pm< 0.01$\\
  NGC0450                 &$2.74\pm0.03$&$1.32\pm0.07$&$3.00\pm< 0.01$&$1.35\pm< 0.01$&$3.07\pm< 0.01$&$1.35\pm< 0.01$&$2.92\pm< 0.01$&$1.42\pm< 0.01$\\
  NGC0941                 &$2.11\pm0.01$&$1.48\pm0.04$&$2.15\pm< 0.01$&$1.71\pm< 0.01$&$2.20\pm< 0.01$&$1.71\pm< 0.01$&$2.17\pm< 0.01$&$1.95\pm< 0.01$\\
  NGC2701                 &$3.71\pm0.04$&$1.20\pm0.01$&$3.53\pm< 0.01$&$1.22\pm< 0.01$&$3.79\pm< 0.01$&$1.20\pm< 0.01$&$3.44\pm< 0.01$&$1.40\pm< 0.01$\\
  NGC2967                 &$2.48\pm0.01$&$5.50\pm0.13$&$2.50\pm< 0.01$&$5.56\pm0.01$&$2.56\pm< 0.01$&$5.77\pm0.02$&$2.49\pm< 0.01$&$5.95\pm0.01$\\
  NGC3055                 &$2.43\pm0.02$&$1.30\pm0.02$&$2.20\pm< 0.01$&$1.18\pm< 0.01$&$2.22\pm< 0.01$&$1.17\pm< 0.01$&$2.24\pm< 0.01$&$1.22\pm< 0.01$\\
  NGC3259                 &$2.06\pm0.01$&$4.83\pm0.09$&$2.07\pm< 0.01$&$4.81\pm0.01$&$2.27\pm< 0.01$&$4.86\pm< 0.01$&$2.07\pm< 0.01$&$5.10\pm0.01$\\
  NGC3423                 &$2.81\pm0.03$&$0.95\pm0.01$&$2.39\pm< 0.01$&$1.01\pm< 0.01$&$2.51\pm< 0.01$&$1.02\pm< 0.01$&$2.35\pm< 0.01$&$1.04\pm< 0.01$\\
  NGC3589                 &$3.39\pm0.04$&$1.59\pm0.02$&$3.06\pm< 0.01$&$1.49\pm< 0.01$&$3.31\pm< 0.01$&$1.39\pm< 0.01$&$3.13\pm< 0.01$&$1.64\pm< 0.01$\\
  NGC3982                 &$1.23\pm0.01$&$1.90\pm0.04$&$1.26\pm< 0.01$&$1.94\pm< 0.01$&$1.27\pm< 0.01$&$2.08\pm< 0.01$&$1.28\pm< 0.01$&$1.90\pm< 0.01$\\
  NGC4041                 &$1.94\pm0.01$&$3.22\pm0.05$&$2.03\pm< 0.01$&$3.93\pm0.01$&$1.98\pm< 0.01$&$4.05\pm0.01$&$2.12\pm< 0.01$&$4.49\pm0.01$\\
  NGC4102                 &$2.76\pm0.06$&$1.19\pm< 0.01$&$2.46\pm< 0.01$&$1.14\pm< 0.01$&$2.36\pm< 0.01$&$1.15\pm< 0.01$&$2.52\pm< 0.01$&$1.16\pm< 0.01$\\
  NGC4545                 &$3.11\pm0.01$&$2.21\pm0.10$&$3.10\pm< 0.01$&$2.16\pm< 0.01$&$3.22\pm< 0.01$&$2.00\pm< 0.01$&$3.12\pm< 0.01$&$2.25\pm< 0.01$\\
  NGC4653                 &$4.34\pm0.02$&$2.74\pm0.14$&$4.39\pm< 0.01$&$3.30\pm< 0.01$&$4.67\pm< 0.01$&$3.10\pm0.01$&$4.34\pm< 0.01$&$3.31\pm0.01$\\
  NGC4904                 &$2.77\pm0.02$&$1.26\pm0.01$&$2.19\pm< 0.01$&$1.20\pm< 0.01$&$2.42\pm< 0.01$&$1.15\pm< 0.01$&$2.09\pm< 0.01$&$1.25\pm< 0.01$\\
  NGC5147                 &$2.34\pm0.03$&$1.22\pm0.01$&$2.01\pm< 0.01$&$1.22\pm< 0.01$&$2.20\pm< 0.01$&$1.18\pm< 0.01$&$1.93\pm< 0.01$&$1.25\pm< 0.01$\\
  NGC5300                 &$3.64\pm0.01$&$1.68\pm0.02$&$3.50\pm< 0.01$&$1.87\pm< 0.01$&$3.83\pm< 0.01$&$1.82\pm0.01$&$3.43\pm< 0.01$&$1.94\pm< 0.01$\\
  NGC5334                 &$4.67\pm0.03$&$2.21\pm0.11$&$4.78\pm< 0.01$&$2.34\pm< 0.01$&$5.00\pm< 0.01$&$2.40\pm< 0.01$&$4.69\pm< 0.01$&$2.57\pm< 0.01$\\
  NGC5376                 &$4.92\pm0.04$&$3.05\pm0.01$&$4.76\pm< 0.01$&$3.11\pm< 0.01$&$4.97\pm0.01$&$3.09\pm< 0.01$&$4.67\pm< 0.01$&$3.22\pm< 0.01$\\
  NGC5430                 &$4.23\pm0.06$&$2.14\pm0.04$&$3.40\pm< 0.01$&$2.31\pm< 0.01$&$3.50\pm< 0.01$&$2.37\pm< 0.01$&$3.37\pm< 0.01$&$2.43\pm< 0.01$\\
  NGC5480                 &$1.33\pm0.01$&$2.91\pm0.14$&$1.22\pm< 0.01$&$1.91\pm0.02$&$1.14\pm< 0.01$&$1.83\pm0.02$&$1.30\pm< 0.01$&$1.65\pm0.03$\\
  NGC5668                 &$4.99\pm0.11$&$3.20\pm0.02$&$4.20\pm< 0.01$&$3.35\pm< 0.01$&$4.18\pm< 0.01$&$3.09\pm< 0.01$&$4.17\pm< 0.01$&$3.71\pm< 0.01$\\
  NGC5693                 &$3.30\pm0.03$&$1.78\pm0.05$&$2.92\pm< 0.01$&$2.02\pm< 0.01$&$3.22\pm< 0.01$&$1.89\pm< 0.01$&$2.81\pm< 0.01$&$2.10\pm< 0.01$\\
  NGC5774                 &$4.25\pm0.06$&$3.08\pm0.05$&$4.40\pm< 0.01$&$3.18\pm0.01$&$4.47\pm0.01$&$3.01\pm0.01$&$4.34\pm< 0.01$&$3.39\pm< 0.01$\\
  NGC7437                 &$3.59\pm0.03$&$2.18\pm0.02$&$3.44\pm< 0.01$&$2.25\pm< 0.01$&$3.36\pm< 0.01$&$2.15\pm< 0.01$&$3.51\pm< 0.01$&$2.36\pm< 0.01$\\
  PGC006667               &$2.94\pm0.03$&$1.48\pm0.07$&$2.80\pm< 0.01$&$1.75\pm0.01$&$2.77\pm< 0.01$&$1.40\pm0.01$&$2.94\pm< 0.01$&$2.04\pm0.01$\\
  UGC02081                &$3.43\pm0.03$&$1.73\pm0.15$&$3.87\pm< 0.01$&$2.92\pm0.01$&$3.98\pm< 0.01$&$2.91\pm0.02$&$3.75\pm< 0.01$&$3.06\pm0.02$\\
  UGC06518                &$1.53\pm0.01$&$1.36\pm0.04$&$1.53\pm< 0.01$&$1.38\pm< 0.01$&$1.54\pm< 0.01$&$1.31\pm< 0.01$&$1.55\pm< 0.01$&$1.45\pm< 0.01$\\
  UGC06903                &$5.13\pm0.09$&$1.50\pm0.04$&$5.46\pm< 0.01$&$1.80\pm< 0.01$&$6.04\pm< 0.01$&$1.68\pm< 0.01$&$5.11\pm< 0.01$&$1.97\pm< 0.01$\\
  UGC07700                &$9.83\pm0.17$&$2.18\pm0.12$&$5.68\pm< 0.01$&$2.92\pm0.01$&$5.54\pm< 0.01$&$2.74\pm< 0.01$&$6.16\pm< 0.01$&$3.43\pm< 0.01$\\
  UGC08084                &$5.92\pm0.11$&$1.63\pm0.09$&$4.28\pm< 0.01$&$2.09\pm0.01$&$4.22\pm< 0.01$&$1.98\pm< 0.01$&$4.44\pm0.01$&$2.40\pm< 0.01$\\
  UGC08658                &$4.10\pm0.03$&$2.92\pm0.03$&$4.07\pm< 0.01$&$3.17\pm< 0.01$&$4.44\pm< 0.01$&$3.24\pm< 0.01$&$3.92\pm< 0.01$&$3.35\pm< 0.01$\\
  UGC09741                &$1.02\pm0.01$&$2.55\pm0.03$&$1.12\pm< 0.01$&$2.48\pm< 0.01$&$0.99\pm< 0.01$&$2.55\pm< 0.01$&$1.21\pm< 0.01$&$2.48\pm< 0.01$\\
  UGC09837                &$3.89\pm0.08$&$1.39\pm0.05$&$3.51\pm< 0.01$&$1.67\pm0.01$&$3.69\pm< 0.01$&$1.52\pm< 0.01$&$3.47\pm< 0.01$&$1.99\pm0.01$\\
  UGC12709                &$5.48\pm0.09$&$2.15\pm0.09$&$4.98\pm0.01$&$2.61\pm< 0.01$&$4.97\pm0.01$&$2.16\pm< 0.01$&$4.98\pm< 0.01$&$2.88\pm0.01$\\
 \hline
\end{tabular}
}
\caption{Derived scale lengths for the ellipse-fit and EP. Units in kpc.
Errors are formal. The slopes of the PAS profiles are los-convlved and therefore not directly translatable into scale lengths, so we have omitted results from that method.}\label{tbl:scalelengths}
\end{table*}

\begin{table*}
\centering
{\small
\begin{tabular}{ l|ll|llllll|llllll }
\textbf{} & \multicolumn{2}{c|}{\textbf{Ellipse}} & \multicolumn{6}{c|}{\textbf{EP}} &  \multicolumn{6}{c}{\textbf{PAS}} \\ 
\textbf{Galaxy} & $r_f$ r' & & $r_f$ r' &&$r_f$ g' &&$r_f$ i' &&$r_f$ r' &&$r_f$ g' &&$r_f$ i'\\ \hline
 IC1125                  &6.978&0.166&6.910&$<0.001$&6.656&0.003&6.147&0.003&7.050&0.066&7.095&0.027&7.484&0.067\\
  IC1158                  &11.575&0.103&11.801&0.006&11.522&0.002&11.849&0.005&11.551&0.041&11.318&0.067&11.687&0.056\\
  NGC0450                 &14.471&0.137&14.077&0.003&13.962&0.002&14.121&0.002&13.535&0.106&13.792&0.041&13.071&0.120\\
  NGC0941                 &13.509&0.157&13.087&0.013&12.749&0.011&14.655&0.014&12.039&0.078&12.126&0.076&11.469&0.145\\
  NGC2701                 &7.322&0.035&7.386&0.004&7.377&0.004&7.159&0.004&6.851&0.062&6.943&0.033&6.800&0.045\\
  NGC2967                 &14.364&0.118&14.128&0.004&14.075&0.010&13.881&0.007&14.927&0.092&14.088&0.092&14.149&0.070\\
  NGC3055                 &9.046&0.082&9.638&$<0.001$&9.470&0.002&9.690&0.001&9.548&0.087&9.801&0.067&9.015&0.070\\
  NGC3259                 &8.762&0.067&8.344&0.003&8.661&0.002&8.264&0.004&7.532&0.105&8.531&0.126&7.844&0.095\\
  NGC3423                 &15.077&0.115&15.462&0.002&15.154&0.008&15.563&0.001&16.490&0.053&15.880&0.043&16.595&0.046\\
  NGC3589                 &6.039&0.053&5.894&0.001&5.850&0.001&5.904&$<0.001$&5.307&0.023&5.283&0.030&4.825&0.059\\
  NGC3982                 &9.260&0.077&9.200&0.006&9.091&0.001&9.271&0.004&7.179&0.074&7.327&0.056&7.673&0.091\\
  NGC4041                 &13.233&0.134&14.211&0.009&14.570&0.006&14.190&0.008&14.644&0.253&15.977&0.334&13.289&0.307\\
  NGC4102                 &9.317&0.059&10.232&$<0.001$&10.378&0.002&10.165&$<0.001$&9.550&0.058&9.549&0.059&9.534&0.049\\
  NGC4545                 &9.708&0.174&9.991&0.001&10.086&0.001&10.130&0.001&9.651&0.037&9.583&0.032&9.920&0.023\\
  NGC4653                 &19.023&0.138&15.169&0.005&15.479&0.022&16.104&0.027&16.377&0.220&16.472&0.035&17.143&0.051\\  
NGC4904                 &7.271&0.043&8.358&0.005&7.938&0.004&8.731&0.011&7.369&0.024&7.106&0.028&7.152&0.032\\
  NGC5147                 &4.909&0.062&5.098&0.001&5.240&$<0.001$&5.191&0.002&5.643&0.035&5.975&0.038&5.875&0.050\\
  NGC5300                 &13.997&0.111&13.205&0.027&12.824&0.042&13.463&0.020&15.143&0.066&14.825&0.087&14.957&0.074\\
  NGC5334                 &16.679&0.259&16.117&0.008&15.778&0.006&16.165&0.006&14.847&0.094&14.308&0.061&15.967&0.052\\
  NGC5376                 &5.054&0.054&4.989&0.001&4.882&0.003&5.010&0.001&5.938&0.032&5.568&0.039&5.947&0.031\\
  NGC5430                 &8.756&0.164&9.778&0.002&9.212&0.012&9.843&0.006&8.795&0.027&8.360&0.050&8.643&0.041\\
  NGC5480                 &10.239&0.213&5.556&0.175&5.061&0.221&-0.523&1.000&4.314&0.146&5.497&0.174&6.705&0.185\\
  NGC5668                 &11.563&0.088&12.335&0.002&12.645&0.003&9.263&0.021&14.602&0.057&13.225&0.112&14.102&0.066\\
  NGC5693                 &6.329&0.157&6.527&0.004&6.224&0.003&6.680&0.002&6.085&0.048&6.488&0.065&5.890&0.102\\
  NGC5774                 &16.470&0.416&16.426&0.005&15.913&0.034&17.501&0.008&13.589&0.057&14.263&0.066&14.282&0.069\\
  NGC7437                 &6.316&0.095&5.973&0.004&5.704&0.013&6.302&0.007&7.311&0.059&7.080&0.119&4.729&0.249\\
  PGC006667               &12.400&0.259&11.955&0.036&12.541&0.036&12.246&0.058&12.875&0.112&12.808&0.102&12.854&0.228\\
  UGC02081                &12.189&0.194&10.943&0.008&10.003&0.037&10.532&0.038&9.891&0.131&8.947&0.193&9.354&0.148\\
  UGC06518                &4.439&0.985&4.806&0.017&4.468&0.001&4.960&0.027&4.198&0.065&4.211&0.022&4.124&0.070\\
  UGC06903                &10.587&0.085&9.965&0.004&9.991&0.002&9.835&0.005&9.816&0.049&10.070&0.060&9.429&0.051\\
  UGC07700                &7.569&0.143&7.398&0.010&7.836&0.005&6.607&0.004&8.284&0.072&8.786&0.072&7.471&0.194\\
  UGC08084                &6.887&0.121&6.762&0.009&6.840&0.004&6.754&0.006&6.868&0.070&5.449&0.095&5.664&0.312\\
  UGC08658                &10.209&0.101&9.605&0.002&9.695&0.003&9.269&0.011&8.928&0.106&9.090&0.092&10.540&0.089\\
  UGC09741                &3.647&0.030&3.681&0.001&3.667&0.001&3.774&0.001&3.433&0.035&3.008&0.039&3.212&0.063\\
  UGC09837                &8.569&0.070&8.032&0.011&8.182&0.005&7.824&0.014&8.532&0.046&8.747&0.046&8.978&0.091\\
  UGC12709                &12.401&0.136&12.104&0.007&11.980&0.005&12.313&0.020&13.096&0.037&12.311&0.136&12.808&0.087\\
\hline
\end{tabular}
}
\caption{Derived break radii. Radii are given in arcsec.}\label{tbl:breakradii}
\end{table*}

{\small
\begin{table*}

\centering
\begin{tabular}{ l|ll|llllll|llllll }
\textbf{} & \multicolumn{2}{c|}{\textbf{Ellipse}} & \multicolumn{6}{c|}{\textbf{EP}} &  \multicolumn{6}{c}{\textbf{PAS}} \\ 
\textbf{Galaxy} &  $\mu_f$ r'& $\pm$ &  $\mu_f$ r' & $\pm$ &  $\mu_f$ g'& $\pm$ &    $\mu_f$ i' & $\pm$ &   $\mu_f$ r' & $\pm$ &   $\mu_f$ g' & $\pm$ &  $\mu_f$ i' & $\pm$ \\ \hline
  IC1125                  &23.613&0.075&23.630&$<0.001$&23.971&0.001&23.025&0.001&18.822&0.077&19.301&0.033&18.978&0.034\\
  IC1158                  &23.980&0.042&24.080&0.002&24.541&0.001&23.728&0.002&18.977&0.028&19.261&0.028&18.528&0.039\\
  NGC0450                 &24.061&0.029&23.899&0.001&24.169&0.001&23.766&$<0.001$&18.118&0.073&18.304&0.004&17.870&0.060\\
  NGC0941                 &24.798&0.074&24.684&0.006&24.911&0.005&24.959&0.004&18.834&0.072&19.440&0.040&18.306&0.046\\
  NGC2701                 &22.246&0.020&22.265&0.002&22.604&0.002&21.956&0.002&16.656&0.016&16.979&0.016&16.482&0.012\\
  NGC2967                 &24.819&0.036&24.663&0.001&25.084&0.003&24.275&0.003&19.341&0.061&19.209&0.045&18.656&0.058\\
  NGC3055                 &22.907&0.044&23.265&$<0.001$&23.613&0.001&23.035&0.001&18.369&0.042&19.047&0.057&17.744&0.058\\
  NGC3259                 &24.295&0.029&23.908&0.001&24.369&0.001&23.608&0.002&18.282&0.022&19.025&0.030&18.068&0.034\\
  NGC3423                 &22.941&0.031&23.066&0.001&23.463&0.002&22.913&$<0.001$&17.314&0.015&17.541&0.021&17.166&0.012\\
  NGC3589                 &23.177&0.029&23.096&$<0.001$&23.357&$<0.001$&22.906&$<0.001$&18.158&0.017&18.429&0.034&17.712&0.029\\
  NGC3982                 &24.042&0.035&23.944&0.003&24.306&0.001&23.683&0.002&17.421&0.026&17.899&0.022&17.269&0.050\\
  NGC4041                 &24.257&0.024&24.437&0.002&25.051&0.001&24.064&0.002&18.151&0.034&18.993&0.057&17.533&0.097\\
  NGC4102                 &21.618&0.015&21.853&$<0.001$&22.592&0.001&21.436&$<0.001$&16.472&0.018&17.140&0.012&16.073&0.018\\
  NGC4545                 &23.526&0.088&23.675&0.001&24.094&$<0.001$&23.482&$<0.001$&18.379&0.029&18.652&0.022&18.232&0.012\\
  NGC4653                 &26.072&0.067&24.659&0.002&25.234&0.009&24.756&0.011&18.999&0.075&19.307&0.012&18.629&0.022\\
  NGC4904                 &22.069&0.022&22.509&0.002&22.824&0.002&22.391&0.005&16.875&0.008&17.288&0.013&16.524&0.010\\
  NGC5147                 &21.211&0.007&21.410&$<0.001$&21.810&$<0.001$&21.259&0.001&16.450&0.019&16.936&0.020&16.369&0.032\\
  NGC5300                 &23.249&0.028&23.044&0.006&23.405&0.008&22.826&0.005&17.945&0.025&18.299&0.006&17.537&0.015\\
  NGC5334                 &23.976&0.051&23.735&0.002&24.139&0.001&23.448&0.001&17.308&0.042&17.501&0.048&17.378&0.011\\
  NGC5376                 &21.286&0.022&21.248&$<0.001$&21.838&0.001&20.895&$<0.001$&16.789&0.009&17.287&0.017&16.388&0.014\\
  NGC5430                 &22.893&0.079&23.359&0.001&23.656&0.005&23.007&0.002&17.348&0.013&17.610&0.016&16.948&0.016\\
  NGC5480                 &24.281&0.120&22.080&0.117&22.308&0.159&&&16.226&0.100&17.649&0.119&17.268&0.080\\
  NGC5668                 &23.244&0.016&23.372&0.000&23.771&0.001&22.580&0.004&17.790&0.027&17.933&0.010&17.447&0.012\\
  NGC5693                 &23.676&0.111&23.634&0.002&23.942&0.001&23.495&0.001&18.168&0.029&18.877&0.043&17.717&0.048\\
  NGC5774                 &24.649&0.118&24.527&0.001&24.812&0.006&24.516&0.002&17.893&0.014&18.524&0.045&17.832&0.020\\
  NGC7437                 &22.899&0.026&22.789&0.001&23.137&0.003&22.649&0.002&17.845&0.041&18.280&0.059&16.789&0.063\\
  PGC006667               &24.563&0.111&24.418&0.013&25.049&0.016&24.195&0.020&19.128&0.048&19.369&0.065&18.765&0.062\\
  UGC02081                &25.691&0.113&25.127&0.003&25.192&0.017&24.760&0.018&19.652&0.113&19.687&0.116&19.201&0.061\\
  UGC06518                &23.739&0.229&23.969&0.016&24.098&0.001&23.833&0.030&19.063&0.067&19.391&0.020&18.626&0.088\\
  UGC06903                &23.853&0.038&23.763&0.001&24.153&0.001&23.503&0.002&18.093&0.003&18.578&0.017&17.708&0.019\\
  UGC07700                &24.121&0.071&24.053&0.004&24.515&0.002&23.615&0.002&19.245&0.021&19.729&0.030&18.862&0.107\\
  UGC08084                &24.052&0.052&24.012&0.005&24.400&0.002&23.737&0.003&18.974&0.031&18.650&0.014&18.213&0.165\\
  UGC08658                &23.930&0.032&23.698&0.000&24.086&0.001&23.388&0.004&18.187&0.043&18.561&0.006&18.408&0.032\\
  UGC09741                &23.260&0.026&23.250&0.000&23.839&0.000&22.982&0.000&18.265&0.026&18.631&0.025&17.850&0.036\\
  UGC09837                &24.615&0.029&24.403&0.005&24.716&0.003&24.135&0.006&19.087&0.020&19.612&0.023&19.308&0.085\\
  UGC12709                &25.177&0.054&25.140&0.002&25.464&0.002&24.967&0.006&20.017&0.034&20.339&0.127&19.809&0.046\\
\hline
\end{tabular}
\caption{Derived break colors. Break colors are given in ABmag/arcsec$^2$.}\label{tbl:colours}
\end{table*}
}

\subsection{Data Reduction}
Originally, we retrieved the SDSS images straight from the SDSS website at 
www.sdss.org.
These came from Data Release 7 \citep{SDSS7}. 
Most of the galaxies are so large, however, that their outskirts  are 
often not covered by the frame and mosaicking would be required.
Instead of manually mosaicking these images, we opted for a different approach.
We used \textsc{Montage}\footnote{\textsc{Montage} is available at 
montage.ipac.caltech.edu/.} \citep{Jacob2010A}, for the retrieval and mosaicking.
In this paper we focus on the $g'$, $r'$ and $i'$ band images.

The following steps were undertaken.
The reference header of the final image was created using \texttt{mHdr}.
Tasks \texttt{mArchiveList} and \texttt{mArchiveExec} were then used in 
sequence to retrieve the $g'$, $r'$ and $i'$ images from SDSS.
The images were then projected to the reference frame using \texttt{mProjExec}.
The overlaps regions between the images were calculated and extracted, 
using \texttt{mOverlaps} and \texttt{mDiffExec}.
With \texttt{mFitExec} the plane fitting coefficients were calculated 
between all frames.
A model of the background was then created using \texttt{mBgModel}. 
We did allow it to fit the slope and set the maximum number of iterations to 
5.
We correct all frames to the common background using \texttt{mBgExec}.
Finally, the images were joined using \texttt{mAdd}.

Each galaxy is thus composed of a set of SDSS frames, which all have a 
background plane subtracted.
We have also run a test wherein only a constant offset correction was 
performed between frames, but in almost all cases, the plane-corrected 
images were superior to the constant offset corrected images.
Only in the case of some large galaxies, such as {NGC\,1042} and 
{NGC\,1068}, did this approach fail and we were forced to remove 
these galaxies from our sample.

The mosaicking of images depends heavily on the correctness of the 
attached world coordinate system in each frame.
The supplied coordinates were correct for all images, except for 
{NGC\,4210}, where we found that stars were duplicated at multiple 
positions in the final mosaics.
We corrected this using the \texttt
{solve-field} tool from the astrometry.net project to verify and 
correct all headers automatically \citep{Lang2010A}. 
This was done directly after downloading the raw SDSS images 
using \texttt{mArchiveExec}.

\subsection{Calibration}
Having created mosaics for the $g'$, $r'$ and $i'$ bands, we 
need to calibrate them to mag/arcsec$^2$. Similar to \citet{Pohlen2004}, 
we use the TsField table files associated with the original observation 
to get the photometric zero point $aa$, the extinction term $kk$ and the 
$airmass$ coefficients.
The surface brightness zero point is calculated as 
\begin{eqnarray}
 \mu_0 =&& -2.5 \times (0.4 \times [aa + kk \times airmass])  \\
&& +2.5 \times \log_{10}(53.907456 \times 0.396^2) \,\,,
\end{eqnarray}
with an exposure time of $53.907456$\,s and an area per pixel of 
$0.396^2$\,arcsec$^2$.
The final surface brightness is then calculated as 
\begin{equation}
\mu = -2.5 \log_{10}{(counts)} + \mu_0\,\,.
\end{equation}

A series of reference stars was then selected in both the calibrated 
image and the mosaic. 
For both images, we measure the magnitudes of these stars.
Using a linear fit to these magnitudes, the mosaic was then adjusted 
to match the calibration.
On average around 15 stars were used, with a matching error below 
0.05 magnitude.

As we noted before in Section \ref{sec:PASmethods}, the PAS method 
has units of mag/arcsec rather than mag/arcsec$^2$, effectively 
making the value dependent on the projected size of the minor 
axis of the galaxy.
We still follow the above calibration strategy for the PAS, 
but in all subsequent plots will add or subtract a linear 
constant term such that the least bright PAS profile (typically 
the $i'$) starts at the same value as the brightest EP profile 
(typically the $g'$).
This is purely meant to guide the eye in direct comparisons 
between the various profiles and should not be seen as the true calibration.

\subsection{Centreing}
Michael Pohlen kindly provided us with the tables from \citet{pt06}. 
We used the values therein to estimate the centre and position 
angle of the images, based on the 25th magnitude ellipse.
The images were rotated to have their major axis aligned with the 
horizontal axis of the image.
Overall this scheme worked well, and only in some cases did we have 
to  tweak the position angle manually to better correspond to the image.

\subsection{Masking}
Foreground stars and background galaxies are a strong contaminant of 
the surface brightness profiles.
\textsc{Sextractor} was used to create an initial set of masks, based 
on the $r'$-band dataset.
For masks outside the galaxy, set by the outer radius of the ellipse-fit 
profiles, we set the masked regions to zero.
Doing that inside the galaxy would create holes in the profile, so a way 
to average over these parts was required.
We therefore use \textsc{Iraf} package \texttt{fixpix} to interpolate 
the good parts of the image into the masked region.
While far from perfect, this is the best solution for inner regions.
If an object has not been fully masked, its unmasked pixels will 
contaminate the interpolation. 
An RGB (red-green-blue) image was therefore created from the three 
bands, and the quality of the masks was inspected.
We tweak the mask by hand and recreate the RGB image. 
This process was repeated until we were satisfied with the result.
In some cases, the contamination is too strong.
We then resort to disabling those quadrants in the minor-axis 
integrated profiles.
The Equivalent Profiles lack such a feature, and in some cases, 
they clearly suffer for it.
For the worst cases, we therefore remove these galaxies from our sample.

As an alternative scheme for future work, it would also have been 
possible to replace the values of the masking with the expected 
values as measured through an initial ellipse fit.
However, the advantage of using our \texttt{fixpix} solution is 
that we make use of the local structure of the galaxy, rather 
than introduce an idealized symmetric version of the galaxy.

\subsection{Background Subtraction}
Background subtraction is a famous problem in SDSS images, 
where due to the storage of numbers as integers one can only 
measure the background using very large samples \citep{pt06}.
We perform a run of \textsc{ellipse} on the data using the 
default parameters but with fixed inclination, centre and position angle.
The background offset is taken as the mean value of all 
results between one and two times the $R_\textrm{outer}$.
Here we use $R_\textrm{outer}$ to denote the outermost 
projected radius of our profile extraction region, which will cover a 
region well beyond the galaxy.
The one-sigma background noise $\sigma$ is taken by measuring the 
standard deviation of all pixels in that same region.
The uncertainty estimation is performed by using the \textsc{Python} 
\texttt{scipy.stats.bayes\_mvs} to perform a Bayesian fit of a normal 
distribution to the background. 
The uncertainty is based on the average confidence limit for the 
mean.\label{sec:uncertainty}

In the online Appendix \ref{sec:onlineplots}, we present RGB images of the 
background, based on the three bands for selected galaxies. 
In regions of the image where an overlap occurs between two 
SDSS frames, there is a better signal to noise ratio due to the 
double observing time spend at those positions.
Since the position of the individual frames that make up an image 
are not identically placed, this leads to locally different colors 
in the RGB images.
This is expected and we therefore do not worry about this.
However, it does imply that the background does not have a constant 
noise level throughout the entire image.
While this is not in itself a bad thing, it is worth keeping in mind 
when examining the profiles with regard to the uncertainty limit.
The mosaicking also introduces (low) correlated noise by regridding 
(averaging) the pixels.
After that, all images were inspected for the flatness of the background; 
we are left with a sample of 57 galaxies with a stable background.

\section{Results}\label{sec:PASanalysis}
\subsection{Classifications}
We classify the profiles by eye for each method. 
While there will always be  some observer bias in the classification, we try 
to minimize this by only distinguishing into three main classes of profiles.
We follow the basic classification scheme proposed by \citet{Erwin2008A}.
The first class, \textbf{Type I}, is used to classify galaxies which exhibit  
a more-or-less constant exponential decay, as known from the work by 
\citet{Patterson1940A}, \citet{Vaucouleurs1948A} and \citet{Freeman1970A}.
The \textbf{Type II} profiles will denote all galaxies, which show a 
downward break/bend in its profile, similar to \citet{Freeman1970A}.
The \textbf{Type III} profiles refers to the so-called to the 
``anti-truncated'' profiles, as first reported by \citet{Erwin2005A}. 
In these profiles, there is a steep decent of the light, followed 
by a less steep descent.
Unlike \cite{pt06} and \citet{Erwin2008A}, we do not sub-classify 
these profiles further. 

There is no fixed criteria to quantify at which point a profile shows 
so many features that it stops being a simple type I profile, and we 
tend to classify more galaxies as type II compared to \citet{pt06}. 
Four galaxies are classified as type I, 40 galaxies are of type II and 
16 are type III.
Note that some galaxies are associated with multiple classifications 
simultaneously.
Overall, the profile classification is the same, independent of the 
type of profile.
Including the mixed types in each category, the ratios are 7\% of the 
galaxies as type I, 70\% as type II and/or 28\% as type III. 
Comparing this with the original classifications by \citet{pt06}, their 
classifications would have been $12\%$ in type I, $65\%$ in type II and 
$35\%$ in type III.
29 galaxies in this sample are classified by us as purely type II galaxies, 
of which four have been classified by \citet{pt06} as type I and five have 
been classified by them as both types II and III.  
In total, we match the complete classification by \citet{pt06} for 40 out 
of 57 galaxies (70\%).

The type III galaxies are an interesting set. 
Most of these galaxies show clear signs of interaction, be it an asymmetric 
disc (e.g. {NGC853}, Figure \ref{fig:NGC853}) or tidal tails (e.g. {NGC3631}, 
Figure \ref{fig:NGC3631}; {NGC3642}, Figure \ref{fig:NGC3642}).
This has already been reported by \citet{kf11}, who noted that many of the 
type III galaxies are mergers.
Table \ref{tbl:classifications} in the online Appendix shows all 
classifications.


\subsection{Measurement of Scale Lengths and Feature Radii}
Measurement of the scale lengths was done in the following way. 
For each galaxy we have seven profiles ($r'$ ellipse, and $g'$, $r'$ and $i'$ 
for both EP and PAS).
We select the most prominent feature in the profile by eye and try to define 
a fit-region on either side of the feature, where the profiles are showing 
more-or-less linear behavior.
For the region before the feature, its outer radii are denoted by $R_1$ and 
$R_2$, while after the feature the radii are $R_3$ and $R_4$.
Table \ref{tbl:classifications}in the online Appendix shows all these radii.

\citet{Knapen1991A} showed that the choice of radii to which a profile is 
fitted, has a strong effect on the derived scale lengths.
To avoid the introduction of a bias, we use the same region for all of our 
profiles.
This can result in a different position for the feature per technique, and 
in some cases we find that the fitting favors different features entirely.
We therefore classify the overall goodness-of-fit for the entire fitting 
result by eye. 
We only focus our statistical analysis on the galaxies having goodness-of-fit 
quality flag 'G' and a purely type II profile. 
This limits our statistics sample to 29 galaxies. 
The scale lengths are listed in Table \ref{tbl:scalelengths}.

We denote the scale lengths measured before the feature as the inner scale 
length $h_0$ and after the feature as outer scale length $h_f$. 
The feature radius is denoted by $R_f$, where any additional subscript will 
be used to refer to the specific method used. The feature surface brightness 
$\mu$ is measured as the surface brightness of the profile at the feature 
radius $R_f$. The radii and surface brightness are listed per galaxy in 
Table \ref{tbl:breakradii}. The colours are listed in table \ref{tbl:colours}.

The following method is used to estimate the errors.
From each fit region, we randomly drop $10\%$ of the points. 
A linear fit is performed to the remaining $90\%$. 
In cases where multiple fit regions are defined, we perform this simultaneous 
in every region.
The resulting fits will be slightly different compared to the original fits.
By proxy, the resulting feature radii and feature surface brightness will also 
be different.
By repeating this process 100 times, a scatter distribution builds up for each 
variable.
We measure the standard deviation from each distribution, and define the error 
as half-width half-maximum by multiplying it by $\sqrt{2 \ln 2}$.
In some cases, we find that not all fits return realistic result.
This occurs in particular in the more noisy regions. 
When this occurs, the error-bars are denoted using arrows in all subsequent 
graphics.

 \subsection{Sub-classification into Breaks and Truncations}
\begin{figure}
 \centering
 \includegraphics[width=0.5\textwidth]{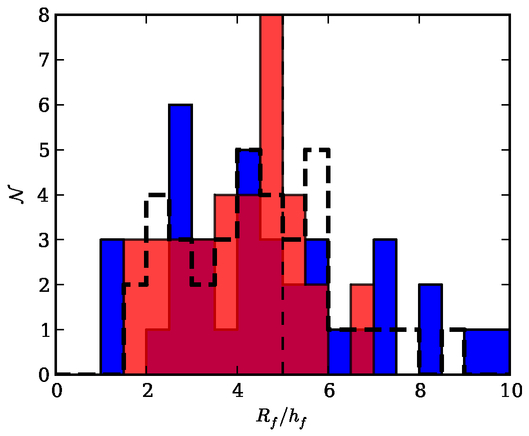}
 \caption[Histogram of $R_f/h_f$ for the final sample]{Histogram of $R_f/h_f$ for the final sample. The dashed vertical line represents the $R_f/h_f=5$ threshold proposed by \citet{mbt12}, with all points to the left considered breaks and all points to the right considered truncations. In light red denotes the EP profiles, while the darker blue denote the PAS profiles. The thick dashed profile shows the distribution of the ellipse-fit profiles.}
 \label{fig:rb_hb}
 \end{figure}

After studying a sample of edge-on galaxies, \citet{mbt12} proposed to 
distinguish between breaks and truncations based on the criteria $R_f/h_f=5$. 
All values above five were considered truncations, while all values lower 
were considered breaks. 
We show the histogram of $R_f/h_f$ for our sample in Figure \ref{fig:rb_hb}.
Twelve truncations are found in the PAS sample.
Seven galaxies in the EP profiles can be considered truncations, out of 
which six overlap with the PAS sample.
We list the truncations sample in Table \ref{tbl:truncations}, along with 
pointers to the figures of the individual galaxies in the online Appendix.
For clarity we repeat that we 
only include here the galaxies having goodness-of-fit 
quality flag 'G' and a purely type II profile.

\begin{figure}
 \centering
 \includegraphics[width=0.5\textwidth]{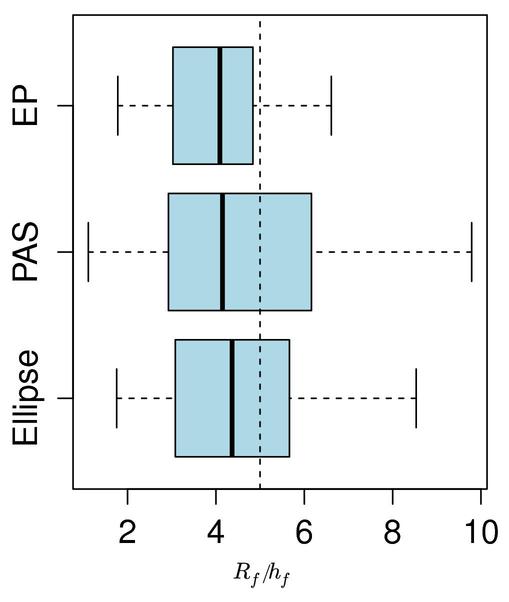}
 \caption[Boxplots of $R_f/h_f$ for the final sample]{Boxplot of $R_f/h_f$ for the final sample. Thick vertical stripes denote the median values, the boxes denote the outlines of the 25\% and 75\%, and the whiskers the minimum and maximum. The dashed vertical line represents the $R_f/h_f=5$ threshold proposed by \citet{mbt12}, with all points to the left considered breaks and all points to the right considered truncations.}
 \label{fig:rb_hbboxplot}
 \end{figure}

Overall, we see that most of the EP $R_f/h_f$ values lie in a compact 
range from 1.5 onto 6. The PAS values span a wider range, between 1.0 
and 10, and are more evenly spread. 
In the edge-on sample from \citet{mbt12} the $R_f/h_f$ ranges from 1 onto 20. 
The mean values of $R_f/h_f$ for the ellipse-fit 
profiles is $4.46\pm2.86$, 
for the EP $4.02\pm1.67$ and for the PAS $4.71\pm5.14$.
The wider spread of the PAS appears to  represent edge-on profiles better 
than the EP profiles.
It is interesting to note that the EP profiles have a lower average value 
and scatter than the ellipse-fit 
profiles. This is due to the redistribution 
of light that occurs as part of the EP method, which has the effect of 
smoothing out the profile a bit.
In Figure \ref{fig:rb_hbboxplot} we show boxplots for all three 
distributions of $R_f/h_f$.
As can be seen more clearly from this figure, the median values of the 
$R_f/h_f$ distributions are similar between the three distributions.
We perform an ANOVA test on the three distributions of $R_f/h_f$, using 
\textsc{R}, to test if the difference in variance is statistically significant. 
We find an F-value of 1.356, which is not statistically significant.

\begin{table}
\centering
\begin{tabular}{lccc}
Galaxy & In EP? & In PAS? & Figure\\
\hline
IC1125 &$\times$ & $\times$ &\ref{fig:IC1125} \\
IC1158 & & $\times$ &\ref{fig:IC1158}  \\
NGC0450& $\times$& $\times$ &\ref{fig:NGC450} \\
NGC0941 & & $\times$ &\\
NGC2701 &  $\times$ & &\\
NGC3055 &$\times$ & $\times$&\ \\
NGC3423 & $\times$& $\times$ &\\
NGC4545 & $\times$& $\times$ &\\
NGC4653 & $\times$& $\times$ &\\
NGC5300 & & $\times$& \\
NGC5430& & $\times$ &\\
PGC006667 & & $\times$ &\\
UGC08084 & & $\times$ &\\
UGC09837 &$\times$ & $\times$ &\\
UGC12709& & $\times$ &\\
\hline
\end{tabular}
\caption[List of truncations]{\label{tbl:truncations}List of truncations based on the \citet{mbt12} criterion}
\end{table}

\subsection{Correlation Tests}
Following \citet{pt06} we perform a range of correlation tests on all our 
parameters. 
We use the Spearman rank correlation coefficient\footnote{We calculate the 
Spearman rank correlation coefficient $\rho$ and significance $p$ using the 
\textsc{Python} package \texttt{scipy.stats.spearmanr}.} $\rho$ to estimate 
if there is any monotone correlation between two parameters. 
We also calculate the corresponding significance $p$ of that $\rho$, using 
as a null-hypothesis the absence of correlation. 
We reject the null-hypothesis when $p<0.05$.
The significance test $p$ only describes the chance of finding a particular 
value of $\rho$ less than or equal to the observed value
purely by chance.
Because of small number statistics, very strong correlations need few samples 
to become statistically significant.
Fainter, but potentially real, correlations require far more samples to 
distinguish from random noise.
As our sample is small, we can only report on relatively strong correlations.

In all subsequent figures, we print the correlation of the combined feature set.
The symbols distinguish between breaks and truncations using filled and 
open markers. The full subset correlation tests can be found in Table 
\ref{tbl:correlations} in the online Appendix.
We will discuss the various correlations in the following subsections. 
Only the most prominent correlations are shown in figures.

\subsubsection{Scale lengths and radii}

 \begin{figure}
 \centering
 \includegraphics[width=0.48\textwidth]{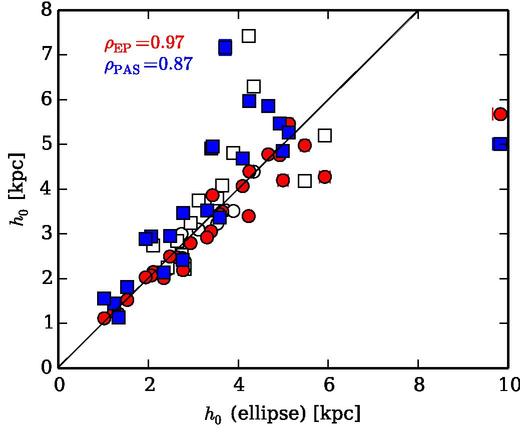}
 \caption[Correlations of inner scale length with method]{The inner scale length of the linear fits to the $r'$-band profile, $h_0$, compared to the scale lengths as derived from the ellipse-fit profiles. The circles denote the EP profiles and the squares denote the PAS profiles. Filled markers are breaks, while open markers are likely truncations. The correlation significance $p$ are also shown.}
 \label{fig:scale0}
 \end{figure}

 \begin{figure}
 \centering
 \includegraphics[width=0.48\textwidth]{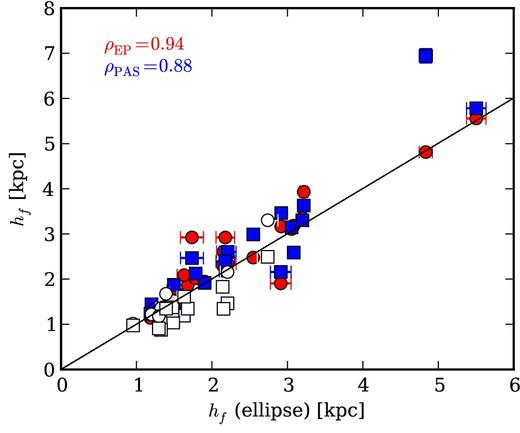}
 \caption[Correlations of outer scale length with method]{The outer scale length of the linear fits to the $r'$-band profile, $h_f$, compared to the scale lengths as derived from the ellipse-fit profiles. The circles denote the EP profiles and the squares denote the PAS profiles. Filled markers are breaks, while open markers are likely truncations. The correlation significance $p$ are also shown.}
 \label{fig:scale1}
 \end{figure}

The $r'$-band inner and outer scale lengths, are shown in Figures 
\ref{fig:scale0} and \ref{fig:scale1}.
From Figure \ref{fig:scale0}, we find that inner scale lengths, as 
measured in both the EP and PAS profiles compared to the ellipse-fit, 
follow very tight positive correlations.
This holds for both the full sample and the sub-samples of truncations 
and breaks. 
The PAS profile have slightly more scatter, and thus the correlations 
are weaker than those of the EP with the ellipse-fit method.
A line is fit through the data, using absolute differences as the cost 
function, and forcing the line through $(0,0)$.
We find that $h_{0,\textrm{EP}}\sim 0.95 h_{0,\textrm{ell}}$ and 
$h_{0,\textrm{PAS}} \sim 1.11 h_{0,\textrm{ell}}$. 
The first relation is likely a result of the inability of the 
EP method to deal with bumps that the ellipse-fit can show. 
If a bump is present in the ellipse-fit profile, this can make the 
scale length in the ellipse-fit profile slightly steeper than that 
of the EP profiles, at least at larger radii than the bump.

Another point to notex
 is more profound: inner scale length of the 
PAS are longer than the profiles in ellipse-fit profiles.
Likely this result is due to the geometry of the galaxies. 
The consequence is that in an edge-on galaxy without dust, the 
inner scale lengths will be longer.
The mean inner scale length $h_0$ for the EP was $3.16 \pm 1.18$\,kpc 
and very similar $3.16 \pm 1.30$\,kpc and $3.17 \pm 0.64$\,kpc for 
the breaks and truncations sub-samples.
The mean inner scale length measured with the PAS is higher and has 
more scatter at $3.87 \pm 1.59$\,kpc, with again rather similar sub-samples 
at $3.81 \pm 1.65$\,kpc and $3.95 \pm 1.49$\,kpc.
The mean inner scale length of the ellipse-fit 
profiles was $3.46\pm1.64$\,kpc.
This is very similar to \citet{pt06}, who found a mean scale length $h_0$ of 
$3.8\pm1.2$\,kpc.

The same tight correlations with the ellipse-fit 
profiles remain for the scale 
lengths after the feature for both profiles.
The biggest scatter increase occurs in the truncations sample of the PAS 
profiles, where $\rho_\textrm{PAS,truncations}=0.732$. 
This result however remains comfortably significant at $p<0.005$. 
The mean outer scale length $h_f$ as measured through the EP profiles  was 
$2.26 \pm 1.03$\,kpc for the EP methods. The breaks have $2.43 \pm 1.04$\,kpc 
and the truncations $1.68 \pm 0.70$\,kpc.
For the PAS we have found $h_f=2.13\pm1.28$\,kpc, with subdivisions into 
$2.61\pm1.41$\,kpc and $1.36\pm0.41$. 
The average ellipse outer scale length $h_f$ is $2.12 \pm 0.98$\,kpc, with 
the breaks at $2.40 \pm 1.08$\,kpc and the truncations at $1.61 \pm 0.47$. 
The outer truncation scale lengths of the EP and ellipse-fit 
profiles thus tend 
to be longer than their PAS counterparts.
In edge-on galaxies, the scale length for a truncation is even shorter, at 
only $1.5\pm0.1$\,kpc.
The break scale lengths is similar at $2.7\pm0.3$\,kpc \citep{mbt12}.

As we noted before, the PAS surface brightness levels are in 
magnitudes / arcsec rather than magnitudes / arcsec$^2$, making 
it distance dependent. 
As a quick test to see if the distance could be a complicating 
fact in the increased scatter of the PAS, we model a very simple 
galaxy consisting of no more than a face-on, truncated exponential disk.
 We have placed at various distances, thus changing 
the number of points available in the minor axis summation. 
In this test, we found that the ratio between the ellipse derived 
scale lengths versus the PAS derived scale lengths did change with 
distance, but the scatter was below $1\%$ of the derived scale lengths. 
Thus, distance does not affect the derived scale lengths significantly.

The feature radii from the ellipse-fit 
profiles to the feature radii measured 
in the other profiles for the r' band images, are also tightly correlated 
between the various methods.
This is expected, as any other result would have caused the profile to be 
flagged.
Again fitting this relation with a linear approximation we find that the 
both the EP and PAS profiles follow $h_{1,\textrm{PAS\&EP}} = 
1.03 h_{f,\textrm{ell}}$.
The mean radius $R_f$ for the EP is $8.38 \pm 2.96$ kpc
in the whole sample. 
The radii $R_f$ for the breaks and truncations sub samples are $8.13 \pm 2.77$ 
and $9.29 \pm 3.39$ kpc.
In comparison, the PAS profile radii $r_f$ have a mean position of 
$8.32\pm3.15$ kpc, 
with the break and truncation sub samples at $7.61 \pm 3.02$ kpc  
and $9.44 \pm 3.03$ kpc. 
The EP results are similar to \citet{pt06}, who report a typical radius of 
$R_f=9.2\pm2.4$ kpc for their type II-CT sample and $R_f=9.5\pm6.5$ for their 
OLR sample.
An average radius of $7.9\pm0.9$ kpc was reported for the inner breaks in the 
edge-on sample of \citet{mbt12}, while the average truncation was found at 
$14\pm2$ kpc.

Looking at the radius of the feature in terms of the number of (inner) scale 
lengths $R_f/h_0$, we have a mean of $2.84 \pm 0.98$ for the EP and 
$2.32 \pm 0.75$ for the PAS.
This difference is a reflection of the result previously reported that PAS 
profiles tend to have longer inner scale lengths compared to the 
ellipse-fit and EP.
The breaks and truncations sub samples for the breaks show similar behavior. 
For the EP we find $2.83 \pm 1.06$ for the breaks and $2.90 \pm 0.67$ for 
the truncations. For the PAS we find $2.20 \pm 0.83$ and $2.52 \pm 0.57$.
Overall, we see that the truncations suffer from less scatter than the breaks.
For truncations in edge-on galaxies, $r_f/h_0$ is expected to lie around 
$4.2\pm0.6$ \citep{vanderKruit1982A} or $2.9\pm0.7$ \citep{2000PDL}.
Sixteen face-on galaxies from the sample of \citet{Wevers1984,Wevers1984A} 
were analyzed 
by \citet{vdk88} for the presence of truncations, who found $r_f/h_0=4.5\pm1.0$.
\citet{Bosma1993} argued that a large range of radii could be found, as seven 
galaxies in the \citet{Wevers1984A} sample, have a relatively bright 'edge' at 
$r_f/h_0=2.8\pm0.4$, while his other did not show this and would thus have 
$R_f/h_0 > 4$. 
\citet{pdla02} find for three galaxies a result of $3.9\pm0.7$.
\citet{pt06} find a far lower $R_f/h_0 \sim 2.5\pm0.6$ for their type II-CT 
sample and $1.7$ for their breaks sample, which we argue 
in more detail in \citet{Peters2015G} is due to different 
definitions of what is a break and truncation plus the way  to mark this.
There also remains the question of the consistency in measuring scale
lengths in face-on galaxies and recovering these from projected
data of edge-ons.
In the literature there are thus studies indicating average values 
around 4 and less than 3, and we see for the moment  no consensus
appearing. The indicative value of 
$3.5-4$ from fig.'s 1 and 2 in \citet{kk04} would seem in the light
of this discussion not to do justice to the small values found and we revise
our indicative value to $3-4$.


\subsubsection{Sharpness of the breaks}\label{sec:h0_hb}
\begin{figure}
 \centering
 \includegraphics[width=0.48\textwidth]{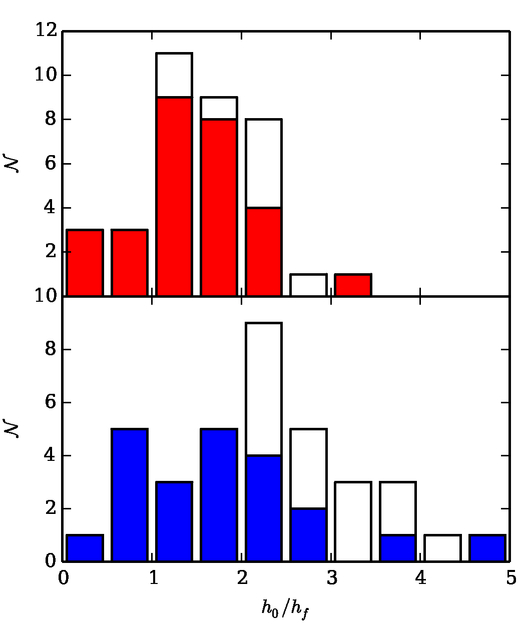}
 \caption[Histogram of $h_0/h_f$ for the final sample]{Histogram of $h_0/h_f$ for the final sample. In the top plot, the results are shown for the EP profile. Filled, red bars are the breaks. Stacked on top of that are the white bars for the truncations. The bottom panel features the results for the PAS profiles, in blue the breaks and in white the truncations.}
 \label{fig:h0_hb}
 \end{figure}

In Figure \ref{fig:h0_hb}, the 'sharpness' of the break is shown as measured 
through the ratio of inner over outer scale lengths $h_0/h_b$.
Breaks from the EP range in strength from just above 0 until 3.5, while the 
PAS range up until 5.
The average EP break sharpness is  $h_0/h_b= 1.57\pm0.64$, while the PAS is 
at  $h_0/h_b= 2.18\pm0.84$.
This reflects the difference between face-on and edge-on profiles, where 
edge-on profiles typically have sharper features, going up to 12 \citep{mbt12}.
Similar to our EP results, \citet{pt06} found $h_0/h_b=2.1\pm0.5$, ranging 
from 1.3 to 3.6, in the type II-CT sample.
Our more extensive classification into type II galaxies creates the 
difference at the lower end of this range.
Truncations and breaks overlap in both profile types, although the 
truncations tend to lie only on higher values.
Compared to the EP, the PAS has a wider and more spread, and can thus 
reach a wider range of sharpness levels.

Figure \ref{fig:h0_hb} offers an interesting alternative to the 
\citet{mbt12} classification scheme of $R_f/h_f=5$. 
As can be seen from the figure, the truncations --as classified 
using the $R_f/h_f=5$ in the PAS-- are all at high values of $h_0/h_f$.
There are however also galaxies that did not meet the $R_f/h_f\geq5$ 
criterion, but which still have high values of $h_0/h_f$. 
An alternative scheme could therefore be to classify all galaxies as 
having a truncation when the sharpness of a break is above some threshold. 
Observing the figure, a value of $h_0/h_b=2$ would seem appropriate,
although we admit thsi has a certain level of arbitrariness.

\begin{figure}
 \centering
 \includegraphics[width=0.48\textwidth]{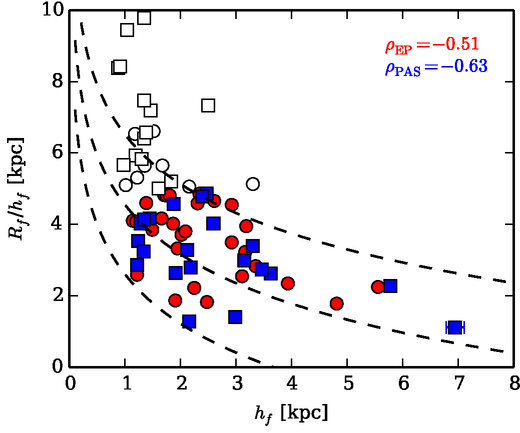}
 \caption[Scale length $h_f$ with ratio $R_f/h_f$]{Scale length $h_f$ with ratio $R_f/h_f$. Filled markers are breaks, open markers are truncations. The top panel shows the results for the PAS and the bottom panel for the EP. Dashed black lines represent the empirical scaling relations from \citet{Schaye2004A}, using $\Sigma_0=5.9$~M$_\odot$~pc$^2$ and $M_\textrm{disc}$ as $5\times10^8$, $3.5\times10^9$ and $25\times10^9$ M$_\odot$.}
 \label{fig:Schaye2004}
 \end{figure}
 
In Figure \ref{fig:Schaye2004}, we plot the relation of $h_f$ with $R_f/h_f$. 
There is a very clear anti-correlation between the two.
\citet{Schaye2004A} predicted the presence of an anti-correlation after 
studying simulations of the thermal and ionization structure of the gaseous 
discs by \citet{Mo1998A}. 
The transition to the cold interstellar medium phase is responsible for the 
onset of local gravitational instability, which triggers star formation.
For an exponential disc, an empirical relation was found which could match 
the data well (Equation \ref{eqn:Schaye2004}). 
Here, $M_\textrm{disc}$ is the total mass of the disc an $\Sigma_c$ is the 
critical face-on surface density of the disc \citep{kk04}, 
\begin{equation}
\frac{R_f}{h_f} = \ln \frac{M_\textrm{disc}}{2\pi h_f^2 \Sigma_c}\,\,.\label{eqn:Schaye2004}
\end{equation}
We apply this model to Figure \ref{fig:Schaye2004}.
The model fits the breaks sub-sample well, but fails to recover the general 
shape of the sample if we include the 
truncations sample.

 
\subsubsection{Variations per band}
 \begin{figure}
 \centering
 \includegraphics[width=0.48\textwidth]{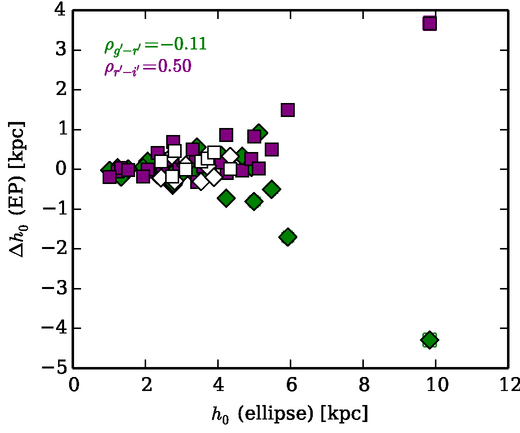}
 \caption[Correlations for differences in scale length $h_0$ per band]{Correlations for differences in scale length $h_0$ per band as measured using the EP profiles, with the scale length from the r'-band ellipse-fit profiles. The corresponding significance values are superimposed on the panel. Diamond shaped boxes represent the $g'\!-\!r'$ points, square boxes represent the $r'\!-\!i'$. Filled markers are breaks, open markers are truncations.}
 \label{fig:scales_colors}
 \end{figure}

It is interesting to test for variations of the scale lengths as measured in 
different bands.
We test this by performing a correlation test of the ellipse-fit 
scale lengths 
$h_0$ and $h_f$ with the differences of the scale lengths measured
in two different bands $\Delta h_0$ and $\Delta h_f$. So e.g.
$\Delta h_0 (g' - r') {\rm (PAS)}$ is  the difference in scale length $h_0$ 
measured in $g'$ and $r'$ using the PAS method.
The results from these tests are shown in Table \ref{tbl:correlations}
(lines 5 to 12).
Only one statistically significant correlation is found. 
The difference of the inner scale lengths measured with the PAS in the $r'$ 
and $i'$ bands has a correlation of $\rho=0.50$ at $p<0.005$. 
We demonstrate this correlation in Figure \ref{fig:scales_colors}. 
The points that lead to this correlation are mostly points with high 
uncertainty, and we thus remain skeptical about any actual correlation.

\begin{table*}
\centering
\begin{tabular}{cc|cc|cc|cc}
Variable 1 & Variable 2 & $\rho$ & $p$ & $\rho_\mathrm{break}$ & $p_\mathrm{break}$ & $\rho_\mathrm{truncation}$ & $p_\mathrm{truncation}$  \\
\hline
$h_0$ (ell) & $h_0$ (EP) & \textbf{0.966} & \textbf{$<$0.01} & \textbf{0.969} & \textbf{$<$0.01} & \textbf{0.952} & \textbf{$<$0.01} \\
$h_0$ (ell) & $h_0$ (PAS) & \textbf{0.874} & \textbf{$<$0.01} & \textbf{0.879} & \textbf{$<$0.01} & \textbf{0.881} & \textbf{$<$0.01} \\
$h_f$ (ell) & $h_f$ (EP) & \textbf{0.942} & \textbf{$<$0.01} & \textbf{0.925} & \textbf{$<$0.01} & \textbf{0.952} & \textbf{$<$0.01} \\
$h_f$ (ell) & $h_f$ (PAS) & \textbf{0.875} & \textbf{$<$0.01} & \textbf{0.927} & \textbf{$<$0.01} & \textbf{0.732} & \textbf{$<$0.01} \\
$h_0$ (ell) & $\Delta h_0$ ($g'-r'$) (PAS) & -0.11 & 0.53 & -0.16 & 0.41 & 0.14 & 0.74 \\
$h_0$ (ell) & $\Delta h_0$ ($r'-i'$) (PAS) & \textbf{0.496} & \textbf{$<$0.01} & \textbf{0.537} & \textbf{$<$0.01} & 0.24 & 0.57 \\
$h_0$ (ell) & $\Delta h_0$ ($g'-r'$) (PAS) & 0.03 & 0.85 & 0.10 & 0.66 & $<$0.01 & 0.99 \\
$h_0$ (ell) & $\Delta h_0$ ($r'-i'$) (PAS) & 0.03 & 0.86 & 0.06 & 0.80 & -0.07 & 0.81 \\
$h_f$ (ell) & $\Delta h_f$ ($g'-r'$) (EP) & 0.22 & 0.20 & 0.19 & 0.33 & 0.33 & 0.42 \\
$h_f$ (ell) & $\Delta h_f$ ($r'-i'$) (EP) & -0.31 & 0.07 & -0.20 & 0.32 & -0.38 & 0.35 \\
$h_f$ (ell) & $\Delta h_f$ ($g'-r'$) (EP) & 0.23 & 0.18 & 0.40 & 0.06 & -0.18 & 0.53 \\
$h_f$ (ell) & $\Delta h_f$ ($r'-i'$) (EP) & -0.15 & 0.38 & -0.18 & 0.43 & 0.29 & 0.31 \\
$R_f$ (ell) & $R_f$ (EP) & \textbf{0.977} & \textbf{$<$0.01} & \textbf{0.973} & \textbf{$<$0.01} & \textbf{0.952} & \textbf{$<$0.01} \\
$R_f$ (ell) & $R_f$ (PAS) & \textbf{0.936} & \textbf{$<$0.01} & \textbf{0.878} & \textbf{$<$0.01} & \textbf{0.960} & \textbf{$<$0.01} \\
$\mu_{r'}$ (EP) & $\mu_{g'-r'}$ (EP) & -0.07 & 0.71 & -0.15 & 0.46 & 0.29 & 0.49 \\
$\mu_{r'}$ (EP) & $\mu_{r'-i'}$ (EP) & -0.05 & 0.77 & -0.02 & 0.92 & -0.24 & 0.57 \\
$\mu_{r'}$ (PAS) & $\mu_{g'-r'}$ (PAS) & -0.26 & 0.13 & -0.34 & 0.13 & 0.17 & 0.55 \\
$\mu_{r'}$ (PAS) & $\mu_{r'-i'}$ (PAS) & 0.30 & 0.08 & \textbf{0.487} & \textbf{0.02} & -0.09 & 0.75 \\
$\mu_{g'-r'}$ (EP) & $v_\textrm{rot}$ & \textbf{0.359} & \textbf{0.03} & 0.37 & 0.05 & 0.19 & 0.65 \\
$\mu_{g'-r'}$ (PAS) & $v_\textrm{rot}$ & 0.19 & 0.25 & 0.27 & 0.23 & 0.13 & 0.65 \\
$\mu_{g'-r'}$ (EP) & $M_\textrm{B,abs}$ & -0.14 & 0.43 & -0.13 & 0.52 & -0.21 & 0.61 \\
$\mu_{g'-r'}$ (PAS) & $M_\textrm{B,abs}$ & -0.08 & 0.64 & -0.13 & 0.56 & -0.06 & 0.84 \\
$\mu_{r'}$ (EP) & $v_\textrm{rot}$ & -0.15 & 0.38 & -0.22 & 0.26 & 0.26 & 0.53 \\
$\mu_{r'}$ (PAS) & $v_\textrm{rot}$ & -0.31 & 0.07 & -0.38 & 0.08 & -0.13 & 0.67 \\
$\mu_{r'}$ (EP) & $M_\textrm{B,abs}$ & 0.09 & 0.58 & 0.12 & 0.55 & 0.33 & 0.42 \\
$\mu_{r'}$ (PAS) & $M_\textrm{B,abs}$ & 0.21 & 0.22 & 0.34 & 0.13 & 0.40 & 0.15 \\
$R_f$ (EP) & $v_\textrm{rot}$ & 0.03 & 0.87 & -0.04 & 0.84 & 0.36 & 0.39 \\
$R_f$ (PAS) & $v_\textrm{rot}$ & 0.04 & 0.82 & -0.02 & 0.93 & 0.34 & 0.24 \\
$R_f$ (EP) & $M_\textrm{B,abs}$ & -0.19 & 0.26 & -0.12 & 0.53 & -0.10 & 0.82 \\
$R_f$ (PAS) & $M_\textrm{B,abs}$ & -0.17 & 0.32 & -0.12 & 0.59 & -0.10 & 0.74 \\
$R_f$ (EP) & $j$ & 0.03 & 0.87 & -0.04 & 0.84 & 0.36 & 0.39 \\
$R_f$ (PAS) & $j$ & 0.04 & 0.82 & -0.02 & 0.93 & 0.34 & 0.24 \\
$a/b$ & $h_0$ (EP) & 0.06 & 0.75 & 0.14 & 0.48 & -0.44 & 0.27 \\
$a/b$ & $h_0$ (PAS) & 0.07 & 0.70 & 0.26 & 0.25 & -0.34 & 0.23 \\
$a/b$ & $h_f$ (EP) & -0.26 & 0.12 & -0.23 & 0.24 & -0.16 & 0.71 \\
$a/b$ & $h_f$ (PAS) & \textbf{-0.404} & \textbf{0.01} & -0.22 & 0.33 & -0.49 & 0.08 \\
$a/b$ & $R_f/h_f$ (EP) & 0.30 & 0.08 & 0.35 & 0.07 & 0.31 & 0.45 \\
$a/b$ & $R_f/h_f$ (PAS) & \textbf{0.349} & \textbf{0.04} & 0.37 & 0.09 & 0.36 & 0.20 \\
$a/b$ & $h_0/h_f$ (EP) & 0.22 & 0.20 & 0.28 & 0.15 & -0.01 & 0.98 \\
$a/b$ & $h_0/h_f$ (PAS) & 0.32 & 0.06 & 0.35 & 0.11 & 0.06 & 0.85 \\
$R_f$ (EP) & $\mu_{g'-r'}$ (EP) & 0.10 & 0.57 & -0.01 & 0.97 & 0.62 & 0.10 \\
$R_f$ (EP) & $\mu_{r'-i'}$ (EP) & 0.21 & 0.23 & 0.30 & 0.13 & -0.12 & 0.78 \\
$R_f$ (PAS) & $\mu_{g'-r'}$ (PAS) & \textbf{-0.349} & \textbf{0.04} & -0.35 & 0.11 & 0.09 & 0.75 \\
$R_f$ (PAS) & $\mu_{r'-i'}$ (PAS) & 0.05 & 0.79 & 0.20 & 0.37 & -0.10 & 0.73 \\
$h_f$ (EP) & $R_f / h_f$ (EP) & \textbf{-0.515} & \textbf{$<$0.01} & \textbf{-0.449} & \textbf{0.02} & -0.17 & 0.69 \\
$h_f$ (PAS) & $R_f / h_f$ (PAS) & \textbf{-0.632} & \textbf{$<$0.01} & \textbf{-0.441} & \textbf{0.04} & -0.37 & 0.19 \\
\hline
\end{tabular}
\caption[Correlation tests for the variables]{Correlation tests for the various variables. Most are self-explanatory, except possibly $j$, which is the specific angular momentum (see text). The $\rho$ stands for the Spearman correlation coefficient. The null-hypothesis is tested with $p$, where values of $p<0.05$ indicate a (2-sigma) significant correlation. These have been highlighted as boldface in the table.}\label{tbl:correlations}
\end{table*}

\begin{table*}
\centering
\begin{tabular}{cccc}
band \& method & full sample & breaks & truncations \\
\hline
$\Delta \mu (g'\!-\!r')$ (EP) & $0.38\pm0.14$ & $0.38\pm0.15$ & $0.39\pm0.09$\\
$\Delta \mu (g'\!-\!r')$ (PAS) & $0.41\pm0.29$ & $0.46\pm0.31$ & $0.32\pm0.23$\\
$\Delta \mu (r'\!-\!i')$ (EP) & $0.25\pm0.18$ & $0.27\pm0.19$ & $0.19\pm0.13$\\
$\Delta \mu (r'\!-\!i')$ (PAS) & $0.29\pm0.34$ & $0.28\pm0.39$ & $0.30\pm0.20$\\\hline
\end{tabular}
\caption{Average brightness differences at the feature radius between various bands}\label{tbl:colordifferences}
\end{table*}

\begin{figure}
 \centering
 \includegraphics[width=0.48\textwidth]{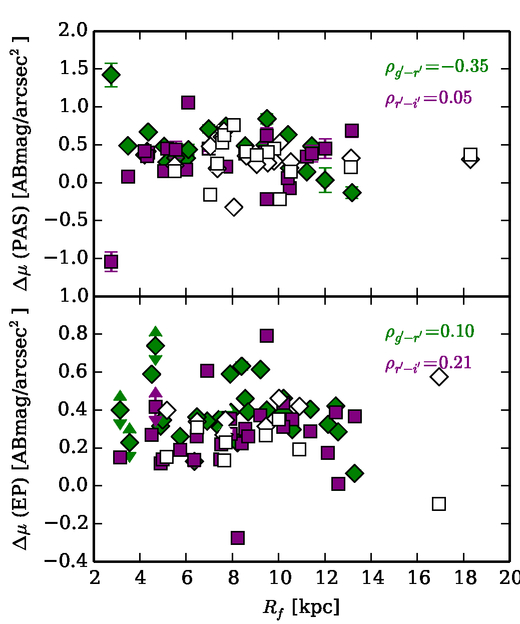}
 \caption[Correlations of the break radius $R_f$ and difference in brightness between bands]{Correlations of the break radius $R_f$, for each method, with the difference in brightness for various bands. Shown are $g'\!-\!r'$ (diamond markers) and $r'\!-\!i'$ (box markers). Filled markers are breaks, open markers are truncations. The top panel shows the results for the PAS and the bottom panel for the EP.}
 \label{fig:radius_color}
 \end{figure}

The differences in brightness between bands at the features from both 
methods are compared in Table \ref{tbl:colordifferences}.
If the error-bars are included in the comparison, the brightness differences 
between bands are practically the same, regardless of the use of either the 
EP or PAS.
The one potential exception to this can be seen in the $r'-i'$ truncations 
sample, where for the EP we have $0.19\pm0.13$ and for the PAS $0.30\pm0.20$. 
As demonstrated earlier, the EP is particularly sensitive to background noise.
Out of all bands, the $i'$ has the highest background noise, so most likely 
the offset is due to sampling of the background. 
We have also tested any potential correlation of feature radius $r_f$ as 
measure by the EP or PAS profiles, with the brightness differences $g'-r'$ 
and $r'-i'$. 
We find a weak correlation of $\rho=-0.35$ and significance $p=0.04$ in the 
PAS $g'-r'$ full-dataset sample, which we show in 
Figure \ref{fig:radius_color}. 
This correlation is likely due to measurements errors of a couple 
of bad points at the outskirts.
 
The r'-band magnitude at the feature has also been compared to the 
$\Delta \mu (r'\!-\!g')$ and $\Delta \mu (r'\!-\!i)'$ values. 
The rank correlation tests show that there is no strong correlation 
for $r'$ with $\Delta \mu (r'\!-\!g')$ and $\Delta \mu (r'\!-\!i')$, 
with the maximum at only $\rho = 0.30$ and $p=0.08$. 
The only exception is for the PAS profiles from the $r'-i'$ breaks 
subsample, where there is a correlation of $\rho=0.49$ and $p=0.02$. 
This correlation is due to the same points as the correlation of $h_0$ 
with $\Delta h_0 (r'-i'0)$ in the PAS.

\subsubsection{Correlations with absolute magnitude, maximum rotation and specific angular momentum}
\citet{mbt12} found that the truncation radius of a galaxy is strongly 
correlated with the maximum rotational velocity $v_\textrm{rot}$ of a 
galaxy, having a correlation of $\rho_\textrm{truncation}=0.81$. 
The breaks in their sample were correlated at only $\rho_\textrm{break}=0.50$.
In contrast, \citet{pt06} report finding no correlation with rotation.
They do report a weak correlation of the brightness at the feature 
$\mu_{r'}$ with the absolute magnitude $M_\textrm{B,abs}$.
We explore possible correlations with of the feature radii $r_f$, 
surface brightness at the feature $\mu$, and differences in brightness between 
bands at the feature $\mu\Delta$, with absolute magnitude $M_\textrm{abs}$, 
maximum rotation $v_\textrm{rot}$, in Table \ref{tbl:correlations}.
We find only one weak correlation in the difference between the $g'$ and 
$r'$ surface brightness at the feature radius, as measured with the EP 
method, with the rotation velocity $v_\textrm{rot}$. 
The correlation strength is $\rho=0.36$ and $p=0.03$.

\citet{mbt12} also perform a correlation test of the break and truncation 
radii with the specific angular momentum $j$, calculated using the empirical 
expression by \citet{Navarro2000}\footnote{In this paper we adopt $h=0.7$},
\begin{equation}
 j \approx 1.3 \times 10^3 \left[ \frac{v_\textrm{rot}}{200 \textrm{km s}^{-1}} \right]^2 \textrm{km s}^{-1} \textrm{h}^{-1} \textrm{kpc}\,\,.\label{eqn:j}
\end{equation}
As this is a rescaling of $v_\textrm{rot}$, the correlation remains the same.
They find that the feature radius only correlates well beyond $r_\textrm{b}=8$ 
kpc. 
For smaller disc with $v_\textrm{rot} < 100 \textrm{km s}^{-1}$, $r_\textrm{break}$ 
and $v_\textrm{rot}$ are essentially unlinked. 
When examining the full radii range, we also do not find a statistically 
significant correlation.
The same holds when limiting ourselves to all features beyond $r>8$.

 \begin{figure*}
 \centering
 \includegraphics[width=0.48\textwidth]{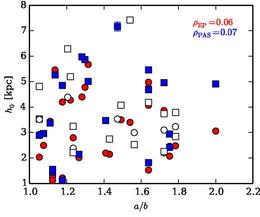}
 \includegraphics[width=0.48\textwidth]{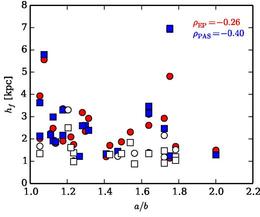}
 \includegraphics[width=0.48\textwidth]{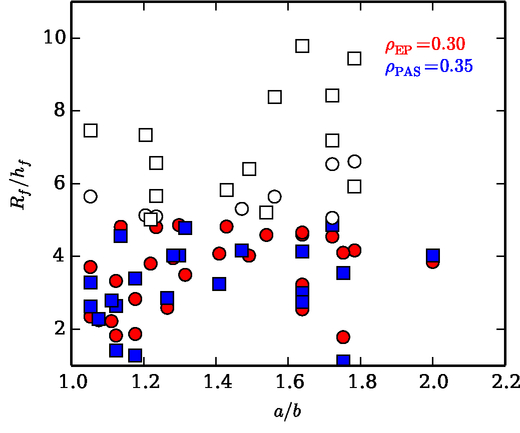}
 \includegraphics[width=0.48\textwidth]{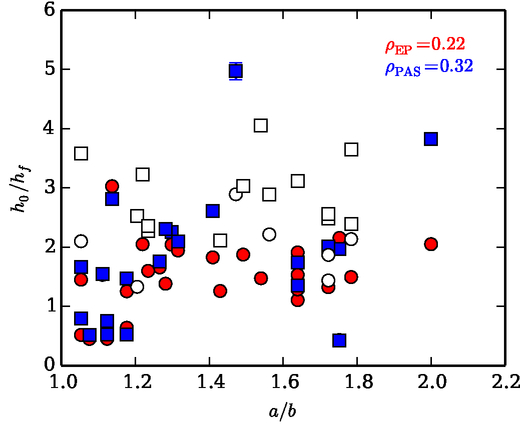}
 \caption[Correlation of various parameters with inclination]{Correlation of various parameters with inclination (expressed as $a/b$). Top row shows with scale length $h_0$ (left) and $h_f$ (right). Bottom row has the $h_0/h_f$ (right) and $R_f/h_f$ (left) correlations. The statistical correlation tests are plotted in each panel. Boxes represent the PAS points, while circles represent the EP points. Open markers represent truncations, while filled markers represent boxes.}
 \label{fig:inclination}
 \end{figure*}

\subsubsection{Effect of inclination}
In Section \ref{sec:h0_hb}, we found that the PAS profiles typically tends to 
have sharper breaks than the ellipse-fit and EP profile.
As the PAS projects data as if it was edge-on, it is also interesting to 
compare the effect of inclinations in this.
To this end, we have tested the correlation of $a/b$ (major axis over minor 
axis) with scale lengths $h_0$ and $h_f$, $h_0/h_f$ and $R_f / h_f$, in 
Table \ref{tbl:correlations}.
The results are also visualized in Figure \ref{fig:inclination}.

We find no significant correlation between $a/b$ and inner scale length $h_0$. 
There may be a negative trend visible in the truncations-only sample, as 
both the EP and PAS profiles have reasonably strong correlations with 
$\rho=-0.44$ and $\rho=-0.34$, but more point will be required before 
this is significant.
When looking at the outer scale length, a negative correlation of $\rho=-0.40$ 
at $p=0.01$ is found for the PAS.
This correlation is mostly due to the truncations samples, which has 
$\rho=-0.49$ at $p=0.08$, which is also clear from the figure. 
Thus, the PAS outer scale length typically gets shorter with higher 
inclination, while the inner scale lengths barely depend on inclination.

The $R_f/h_f$ has a weak but significant correlation with $a/b$ in the 
PAS method, $\rho=0.35$ at $p=0.04$, a correlation which holds (albeit 
insignificantly) for both subsets.
Interestingly the EP results follow similar correlations, although those 
are not significant.
The sharpness of the breaks $h_0/h_f$ has no significant correlation with 
$a/b$ in either method.

\section{Closing Discussion and Conclusions}\label{sec:PASdiscussion}
We have developed two new approaches for extracting the surface photometry of 
a face-on galaxy.
The Equivalent Profiles (EP) work under the assumption that the surface 
brightness of a galaxy decreases as the radius increases. 
By starting  with the brightest pixel and moving to lower brightness levels, 
each level can be assigned an equivalent area ellipse containing the surface 
of all pixels that are at or brighter than that level. The equivalent ellipse 
then gives the equivalent radius.
The other method is the Principle Axis Summation (PAS), which work by summing 
the light onto the principle axis of the galaxy. 
This method then gives the equivalent of the profile as if the galaxy was seen 
edge-on.
We have then tested this method on a sub-sample of the galaxies from 
\citet{pt06}.

Seen overall, we find that both our methods perform well. 
Considering the fundamentally different method used to derive them, 
a detailed comparison that we have made (not illustrated) shows to us 
that the EP are remarkably similar to the ellipse-fit 
profiles as measured by \citet{pt06}. We also point out that the classical
method of elliptical averaging compares very well with results of
equivalent profiles, \citep{vdk79}.
There are some differences. 
The ellipse-fit 
profiles have the ability to measure local upturns in the 
profiles, for example due to a local bar or ring feature.
By design, the EP is unable to cope with this. 
This can lead to slightly different scale lengths.
Beyond such a bump however, the EP profile and ellipse-fit join up again, 
as for example in Figure \ref{fig:IC1067}.
Overall, we see that the EP behaves worse at lower brightness levels 
than the ellipse-fit profiles.
For practical purposes, the ellipse therefore remains the preferred method.

The PAS method turns out to be a very interesting approach. 
Compared to the EP profiles, breaks and truncations often look sharper. 
A good example of this can be seen in galaxy IC1158, seen in Figure 
\ref{fig:IC1158}, where the PAS profiles starts to drop quite rapidly 
beyond $\sim65''$, much stronger than the ellipse-fit profiles.
We find that the inner scale length as measured with the PAS is on average 
10\% longer than the same scale length in either of the other methods. 
We also find a negative correlation with the inclination as expressed by 
the ratio $a/b$. 
As the inclination increase, the outer scale length of the PAS profiles 
get smaller. 
This leads to sharper breaks $h_0/h_f$ than is seen in ellipse-fit 
profiles 
or in the EP profiles. 
Although beyond the scope of this project, it would be interesting to 
test if $h_0/h_f$, rather than $R_b/h_f$ is a good way to distinguish 
between breaks and truncations.

In edge-on galaxies, there is a well-observed correlation of the radius 
of the truncations with the maximum rotation velocity $v_\mathrm{rot}$ 
\citep{vanderKruit2008A}.
This was confirmed by \citet{mbt12}, who also found a correlation with 
the absolute magnitude of the galaxy $M_\textrm{B,abs}$.
Various studies of face-on samples, starting with \citet{pt06} 
and more recently for example \citep{2013mm} have looked for and 
sometimes reported similar relations (e.g. their fig.~8), 
but in those cases it is not clear that
the break radii used are referring to the equivalent features of 
edge-on truncations. 
We do not find any correlation of the surface brightness at the feature, 
difference in brightness between various bands at the feature, and the 
feature radius, with the absolute brightness $M_\textrm{B,abs}$ nor with the 
maximum rotation $v_\textrm{rot}$.
\citet{mbt12} divide their samples up into truncations and breaks based 
on the criteria $R_f/h_0=5$, with the galaxy belonging to breaks if the 
ratio was below five and truncations if it was above it.
We have split our sample into these two subsets using the same criteria and 
have inspected the data for correlations.
We do not reproduce these correlations. 
We are therefore skeptical of the galaxies in our truncations subsample 
constituting true truncations in the edge-on sense. 
It more likely that we are still only observing breaks. 
Truncations can likely only be found by using deeper imaging, such as that 
used by \citet{Bakos2012a}.
We will explore the use of deeper imaging to detect truncations further in 
\citet{Peters2015G}.

\section*{Acknowledgments}
We thank Michael Pohlen for providing his data on the sample in electronic
form.

SPCP is grateful to the Space Telescope Science Institute, Baltimore, USA, the 
Research School for Astronomy and Astrophysics, Australian National University, 
Canberra, Australia, and the Instituto de Astrofisica de Canarias, La Laguna, 
Tenerife, Spain, for hospitality and support during  short and extended
working visits in the course of his PhD thesis research. He thanks
Roelof de Jong and Ron Allen for help and support during an earlier 
period as visiting student at Johns Hopkins University and 
the Physics and Astronomy Department, Krieger School of Arts and Sciences 
for this appointment.

PCK thanks the directors of these same institutions and his local hosts
Ron Allen, Ken Freeman and Johan Knapen for hospitality and support
during many work visits over the years, of which most were 
directly or indirectly related to the research presented in this series op 
papers.

Work visits by SPCP and PCK have been supported by an annual grant 
from the Faculty of Mathematics and Natural Sciences of 
the University of Groningen to PCK accompanying of his distinguished Jacobus 
C. Kapteyn professorhip and by the Leids Kerkhoven-Bosscha Fonds. PCK's work
visits were also supported by an annual grant from the Area  of Exact 
Sciences of the Netherlands Organisation for Scientific Research (NWO) in 
compensation for his membership of its Board.


\bibliography{refsVI}
\bibliographystyle{mn2e}



\appendix

\section{Plots for Individual Galaxies}\label{sec:onlineplots}
The figures in this Appendix show the surface photometry for all galaxies 
in our sample.

The top-left image demonstrates the inner parts of the galaxy. The red 
ellipse denotes the trust radius $R_\textrm{trust}$ in which we have derived 
the PAS profiles.
The white ellipse shows the area in which we have measured the EP profiles.

The top-right image shows an image of the deep background. 
The region between the two red ellipses (one and two times the trust radius 
$R_\textrm{trust}$) is used to estimate the background level. The inner ellipse
is the same as the red one in the upper image.

The bottom panel shows all profiles for the $g'$, $r'$ and $i'$ bands. 
The top set of curves are based on the PAS photometry. From top to bottom 
the set of curves is for the $i'$, $r'$ and $g'$ bands.
The horizontal axis in the profiles
is in arcsec, the vertical one in magnitudes
per arcsecond$^2$.
Note that the PAS has units of mag/arcsec. We have chosen the vertical offset 
such that the reader can easily compare with the other parameters. 
The bottom set of curves are from the EP photometry. The black curves are the 
$r'$ ellipse-fit profiles from \citet{pt06}.
The vertical dashed lines represent the outermost radii for the PAS and EP 
ellipses.
The horizontal dashed lines show the one-sigma noise levels of the background.
\newpage
\begin{figure}
\centering
   \includegraphics[angle=0,width=0.48\textwidth]{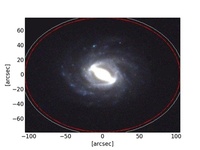}
   \includegraphics[angle=0,width=0.48\textwidth]{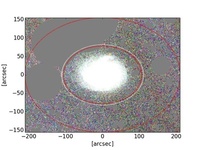}
   \includegraphics[angle=0,width=0.48\textwidth]{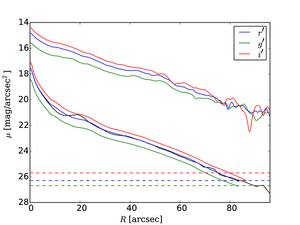}
  \caption{\label{fig:IC1067}IC 1067: PAS is using the PAS is using the lower right and upper right quadrants.}
\end{figure}

\begin{figure}
\centering
   \includegraphics[angle=0,width=0.48\textwidth]{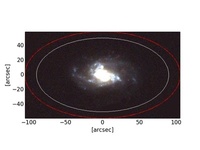}
   \includegraphics[angle=0,width=0.48\textwidth]{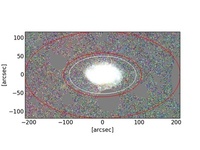}
   \includegraphics[angle=0,width=0.48\textwidth]{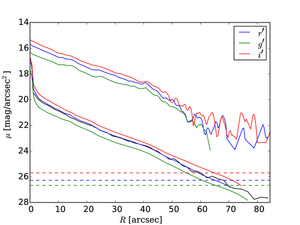}
  \caption{\label{fig:IC1125}IC 1125: PAS is using all quadrants.
}
\end{figure}

\begin{figure}
\centering
   \includegraphics[angle=0,width=0.48\textwidth]{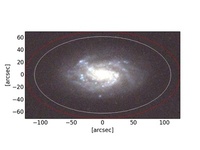}
   \includegraphics[angle=0,width=0.48\textwidth]{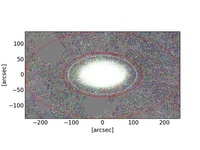}
   \includegraphics[angle=0,width=0.48\textwidth]{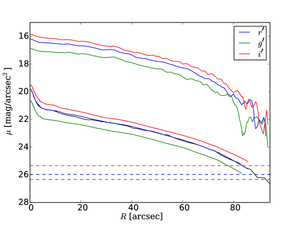}
  \caption{\label{fig:IC1158}IC 1158: PAS is using all quadrants.}
\end{figure}

\begin{figure}
\centering
   \includegraphics[angle=0,width=0.48\textwidth]{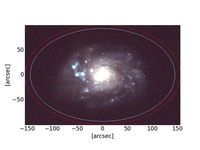}
   \includegraphics[angle=0,width=0.48\textwidth]{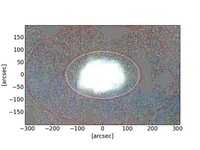}
   \includegraphics[angle=0,width=0.48\textwidth]{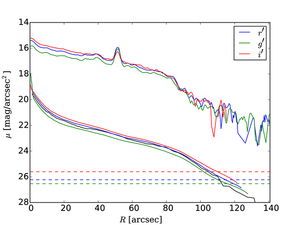}
  \caption{\label{fig:NGC450}NGC 0450: PAS is using the lower left, lower right and upper right quadrants.}
\end{figure}

\begin{figure}
\centering
   \includegraphics[angle=0,width=0.48\textwidth]{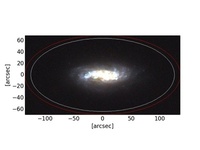}
   \includegraphics[angle=0,width=0.48\textwidth]{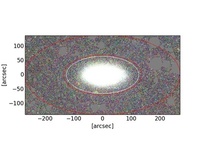}
   \includegraphics[angle=0,width=0.48\textwidth]{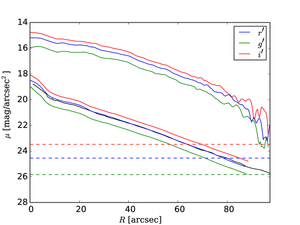}
  \caption{NGC 0701: PAS is using all quadrants. }
\end{figure}

\begin{figure}
\centering
   \includegraphics[angle=0,width=0.48\textwidth]{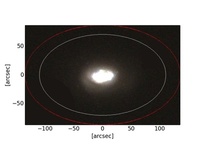}
   \includegraphics[angle=0,width=0.48\textwidth]{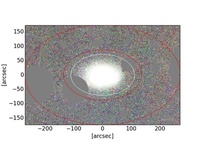}
   \includegraphics[angle=0,width=0.48\textwidth]{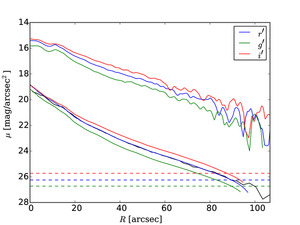}
  \caption{\label{fig:NGC853}NGC 0853: PAS is using the PAS is using the lower left and upper right quadrants. }
\end{figure}

\begin{figure}
\centering
   \includegraphics[angle=0,width=0.48\textwidth]{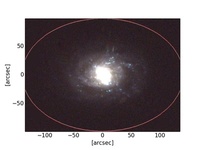}
   \includegraphics[angle=0,width=0.48\textwidth]{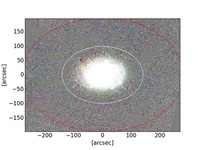}
   \includegraphics[angle=0,width=0.48\textwidth]{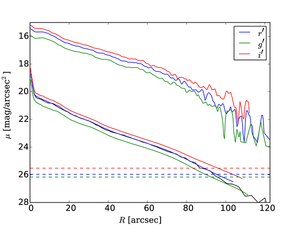}
  \caption{\label{fig:NGC941}NGC 0941: PAS is using all quadrants.}
\end{figure}

\begin{figure}
\centering
   \includegraphics[angle=0,width=0.48\textwidth]{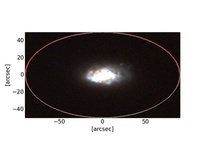}
   \includegraphics[angle=0,width=0.48\textwidth]{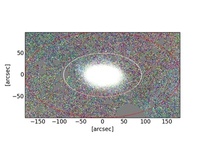}
   \includegraphics[angle=0,width=0.48\textwidth]{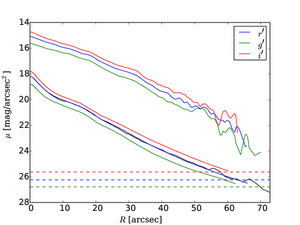}
  \caption{NGC 1299: PAS is using all quadrants.}
\end{figure}

\begin{figure}
\centering
   \includegraphics[angle=0,width=0.48\textwidth]{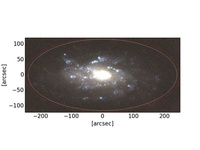}
   \includegraphics[angle=0,width=0.48\textwidth]{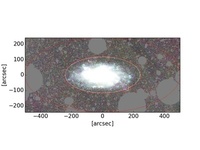}
   \includegraphics[angle=0,width=0.48\textwidth]{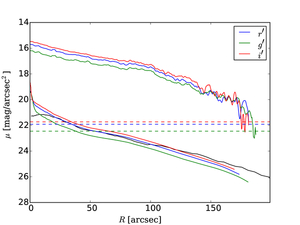}
  \caption{\label{fig:NGC 2541}NGC 2541: is not using the lower right quadrant.}
\end{figure}

\begin{figure}
\centering
   \includegraphics[angle=0,width=0.48\textwidth]{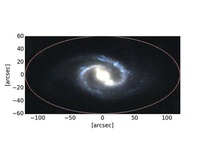}
   \includegraphics[angle=0,width=0.48\textwidth]{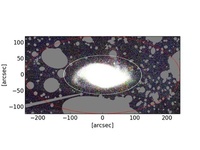}
   \includegraphics[angle=0,width=0.48\textwidth]{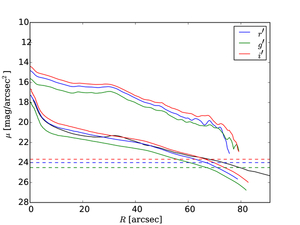}
  \caption{NGC 2543: PAS is using all quadrants.}
\end{figure}
 
\begin{figure}
\centering
   \includegraphics[angle=0,width=0.48\textwidth]{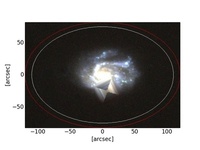}
   \includegraphics[angle=0,width=0.48\textwidth]{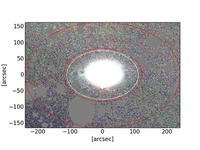}
   \includegraphics[angle=0,width=0.48\textwidth]{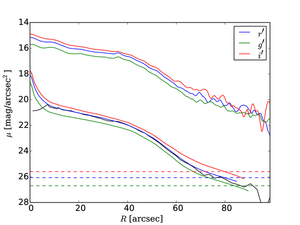}
  \caption{\label{fig:NGC2701}NGC 2701: PAS is using the PAS is using the lower left and lower right quadrants.}
\end{figure}

\begin{figure}
\centering
   \includegraphics[angle=0,width=0.48\textwidth]{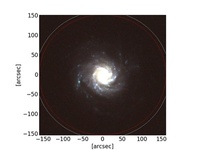}
   \includegraphics[angle=0,width=0.48\textwidth]{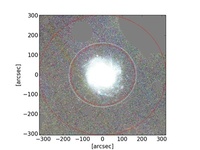}
   \includegraphics[angle=0,width=0.48\textwidth]{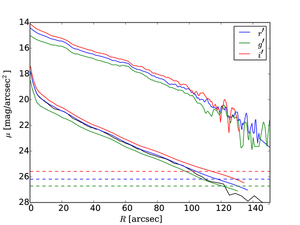}
  \caption{NGC 2776: PAS is using all quadrants.}
\end{figure}

\begin{figure}
\centering
   \includegraphics[angle=0,width=0.48\textwidth]{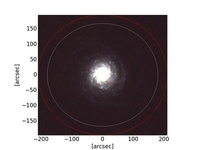}
   \includegraphics[angle=0,width=0.48\textwidth]{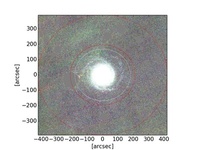}
   \includegraphics[angle=0,width=0.48\textwidth]{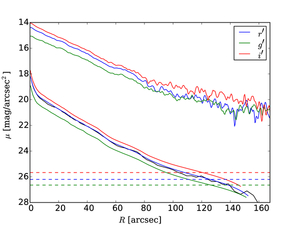}
  \caption{NGC 2967: PAS is using all quadrants.}
\end{figure}

\clearpage 

\begin{figure}
\centering
   \includegraphics[angle=0,width=0.48\textwidth]{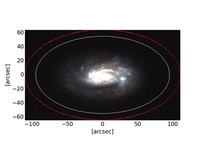}
   \includegraphics[angle=0,width=0.48\textwidth]{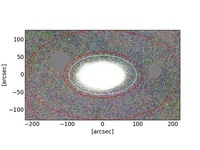}
   \includegraphics[angle=0,width=0.48\textwidth]{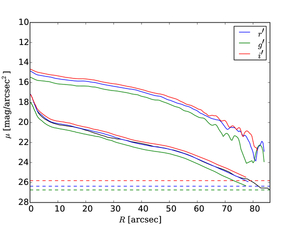}
  \caption{\label{fig:NGC3055}NGC 3055: PAS is using all quadrants.}
\end{figure}

\begin{figure}
\centering
   \includegraphics[angle=0,width=0.48\textwidth]{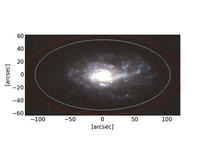}
   \includegraphics[angle=0,width=0.48\textwidth]{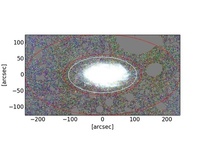}
   \includegraphics[angle=0,width=0.48\textwidth]{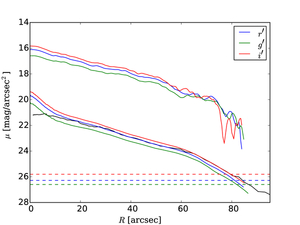}
  \caption{NGC 3246: PAS is using all quadrants.}
\end{figure}

\begin{figure}
\centering
   \includegraphics[angle=0,width=0.48\textwidth]{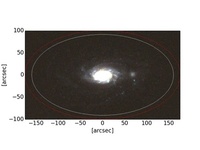}
   \includegraphics[angle=0,width=0.48\textwidth]{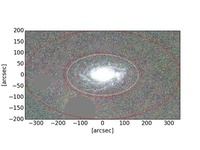}
   \includegraphics[angle=0,width=0.48\textwidth]{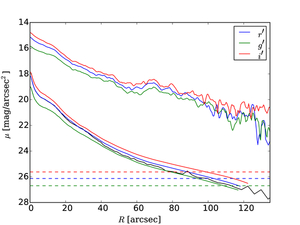}
  \caption{NGC 3259: PAS is using all quadrants.}
\end{figure}

\begin{figure}
\centering
   \includegraphics[angle=0,width=0.48\textwidth]{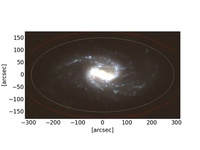}
   \includegraphics[angle=0,width=0.48\textwidth]{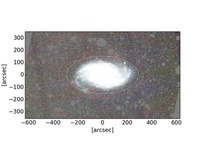}
   \includegraphics[angle=0,width=0.48\textwidth]{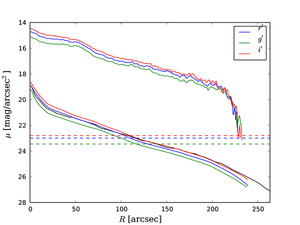}
  \caption{NGC 3359: PAS is using all quadrants.}
\end{figure}

\begin{figure}
\centering
   \includegraphics[angle=0,width=0.48\textwidth]{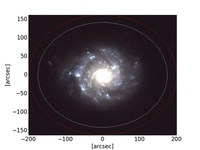}
   \includegraphics[angle=0,width=0.48\textwidth]{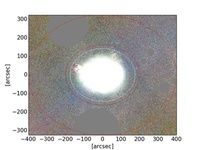}
   \includegraphics[angle=0,width=0.48\textwidth]{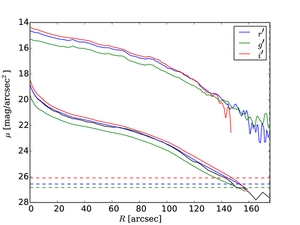}
  \caption{NGC 3423: PAS is using the PAS is using all quadrants.}
\end{figure}

\begin{figure}
\centering
   \includegraphics[angle=0,width=0.48\textwidth]{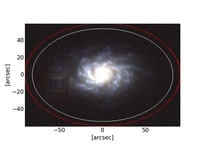}
   \includegraphics[angle=0,width=0.48\textwidth]{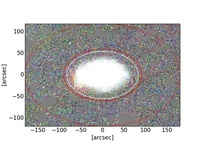}
   \includegraphics[angle=0,width=0.48\textwidth]{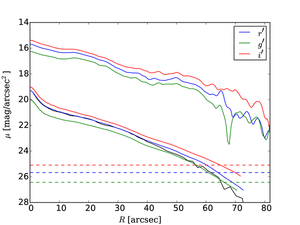}
  \caption{NGC 3488: PAS is using the PAS is using the lower right and upper right quadrants.}
\end{figure}

\begin{figure}
\centering
   \includegraphics[angle=0,width=0.48\textwidth]{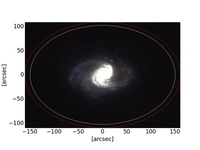}
   \includegraphics[angle=0,width=0.48\textwidth]{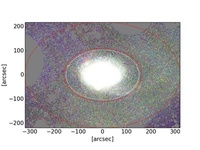}
   \includegraphics[angle=0,width=0.48\textwidth]{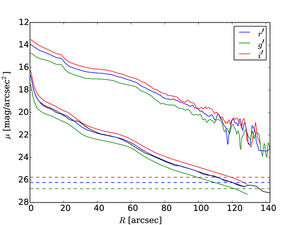}
  \caption{NGC 3583: PAS is using all quadrants.}
\end{figure}

\begin{figure}
\centering
   \includegraphics[angle=0,width=0.48\textwidth]{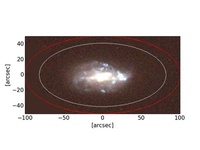}
   \includegraphics[angle=0,width=0.48\textwidth]{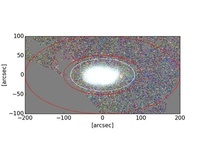}
   \includegraphics[angle=0,width=0.48\textwidth]{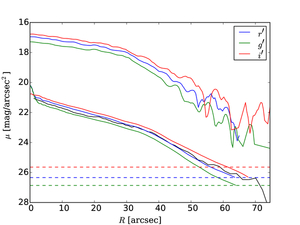}
  \caption{NGC 3589: PAS is using all quadrants.}
\end{figure}

\begin{figure}
\centering
   \includegraphics[angle=0,width=0.48\textwidth]{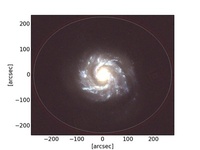}
   \includegraphics[angle=0,width=0.48\textwidth]{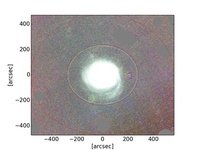}
   \includegraphics[angle=0,width=0.48\textwidth]{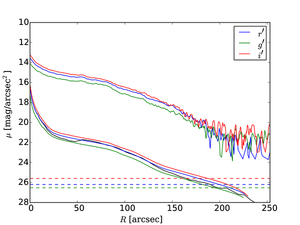}
  \caption{\label{fig:NGC3631}NGC 3631: PAS is using all quadrants.}
\end{figure}

\begin{figure}
\centering
   \includegraphics[angle=0,width=0.48\textwidth]{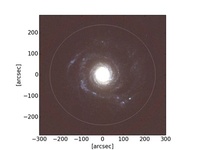}
   \includegraphics[angle=0,width=0.48\textwidth]{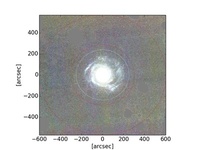}
   \includegraphics[angle=0,width=0.48\textwidth]{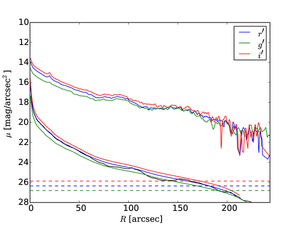}
  \caption{\label{fig:NGC3642}NGC 3642: PAS is using all quadrants.}
\end{figure}

\begin{figure}
\centering
   \includegraphics[angle=0,width=0.48\textwidth]{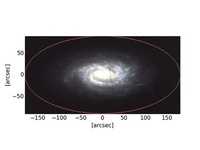}
   \includegraphics[angle=0,width=0.48\textwidth]{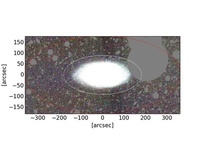}
   \includegraphics[angle=0,width=0.48\textwidth]{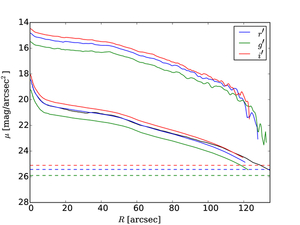}
  \caption{NGC 3756: PAS is not using the top left quadrants.}
\end{figure}

\begin{figure}
\centering
   \includegraphics[angle=0,width=0.48\textwidth]{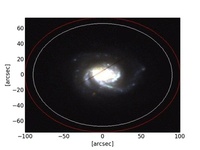}
   \includegraphics[angle=0,width=0.48\textwidth]{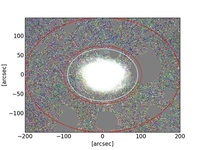}
   \includegraphics[angle=0,width=0.48\textwidth]{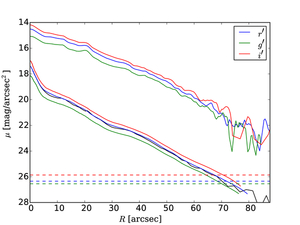}
  \caption{NGC 3888: PAS is using all quadrants.}
\end{figure}

\clearpage

\begin{figure}
\centering
   \includegraphics[angle=0,width=0.48\textwidth]{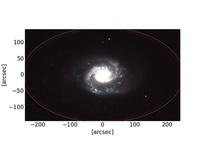}
   \includegraphics[angle=0,width=0.48\textwidth]{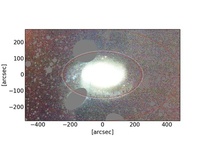}
   \includegraphics[angle=0,width=0.48\textwidth]{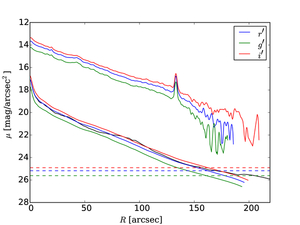}
  \caption{NGC 3893: PAS is using all quadrants.}
\end{figure}

\begin{figure}
\centering
   \includegraphics[angle=0,width=0.48\textwidth]{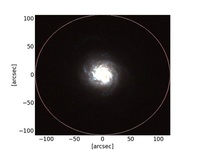}
   \includegraphics[angle=0,width=0.48\textwidth]{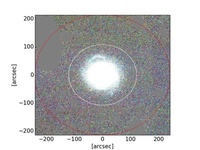}
   \includegraphics[angle=0,width=0.48\textwidth]{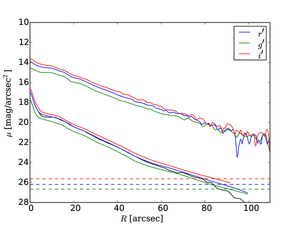}
  \caption{NGC 3982: PAS is using all quadrants.}
\end{figure}
 
\begin{figure}
\centering
   \includegraphics[angle=0,width=0.48\textwidth]{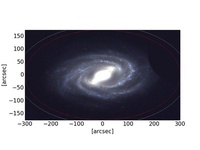}
   \includegraphics[angle=0,width=0.48\textwidth]{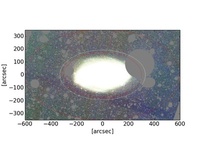}
   \includegraphics[angle=0,width=0.48\textwidth]{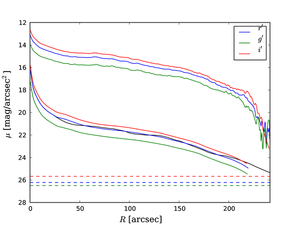}
  \caption{NGC 3992: PAS is using all quadrants.}
\end{figure}

\begin{figure}
\centering
   \includegraphics[angle=0,width=0.48\textwidth]{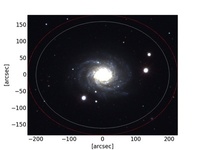}
   \includegraphics[angle=0,width=0.48\textwidth]{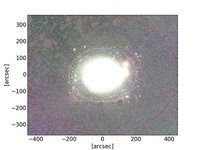}
   \includegraphics[angle=0,width=0.48\textwidth]{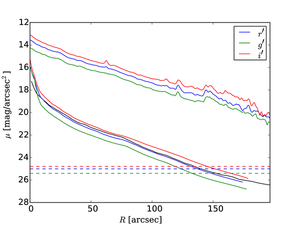}
  \caption{NGC 4030: PAS is using all quadrants.}
\end{figure}

\clearpage

\begin{figure}
\centering
   \includegraphics[angle=0,width=0.48\textwidth]{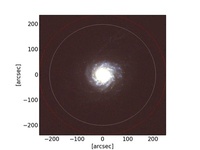}
   \includegraphics[angle=0,width=0.48\textwidth]{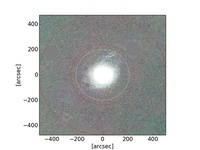}
   \includegraphics[angle=0,width=0.48\textwidth]{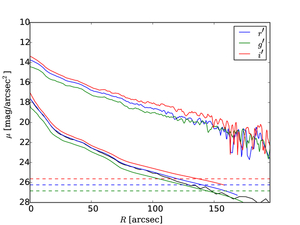}
  \caption{NGC 4041: PAS is using all quadrants.}
\end{figure}

\begin{figure}
\centering
   \includegraphics[angle=0,width=0.48\textwidth]{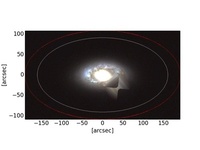}
   \includegraphics[angle=0,width=0.48\textwidth]{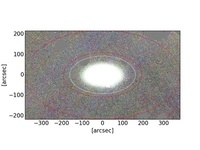}
   \includegraphics[angle=0,width=0.48\textwidth]{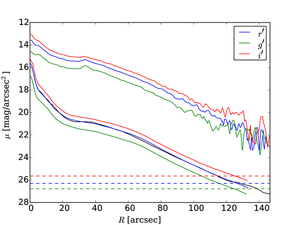}
  \caption{NGC 4102: PAS is using the lower left, upper left and lower right quadrants.}
\end{figure}

\begin{figure}
\centering
   \includegraphics[angle=0,width=0.48\textwidth]{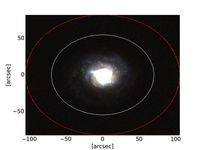}
   \includegraphics[angle=0,width=0.48\textwidth]{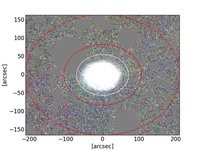}
   \includegraphics[angle=0,width=0.48\textwidth]{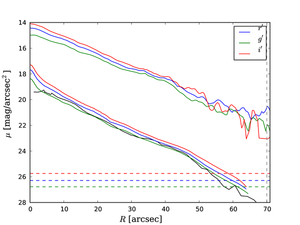}
  \caption{NGC 4108: PAS is using the upper left, lower right and upper right quadrants.}
\end{figure}

\begin{figure}
\centering
   \includegraphics[angle=0,width=0.48\textwidth]{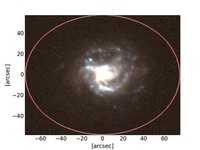}
   \includegraphics[angle=0,width=0.48\textwidth]{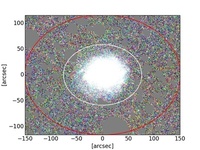}
   \includegraphics[angle=0,width=0.48\textwidth]{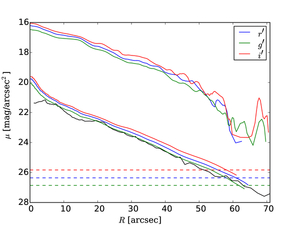}
  \caption{NGC 4108B: PAS is using all quadrants.}
\end{figure}

\begin{figure}
\centering
   \includegraphics[angle=0,width=0.48\textwidth]{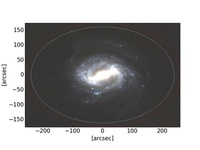}
   \includegraphics[angle=0,width=0.48\textwidth]{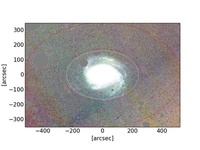}
   \includegraphics[angle=0,width=0.48\textwidth]{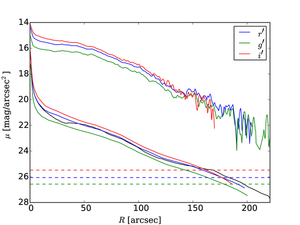}
  \caption{NGC 4123: PAS is using all quadrants.}
\end{figure}

\begin{figure}
\centering
   \includegraphics[angle=0,width=0.48\textwidth]{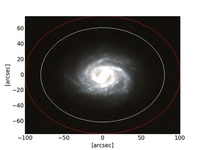}
   \includegraphics[angle=0,width=0.48\textwidth]{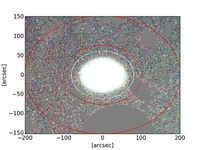}
   \includegraphics[angle=0,width=0.48\textwidth]{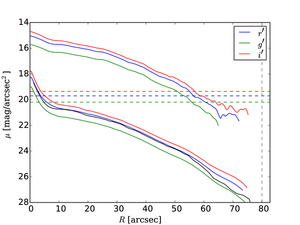}
  \caption{NGC 4210: PAS is using all quadrants.}
\end{figure}

\begin{figure}
\centering
   \includegraphics[angle=0,width=0.48\textwidth]{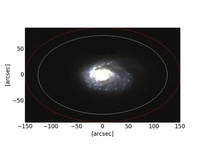}
   \includegraphics[angle=0,width=0.48\textwidth]{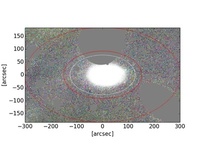}
   \includegraphics[angle=0,width=0.48\textwidth]{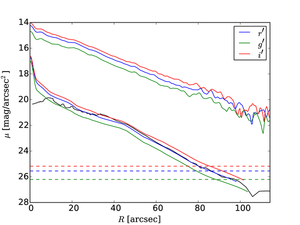}
  \caption{NGC 4273: PAS is using all quadrants.}
\end{figure}

\begin{figure}
\centering
   \includegraphics[angle=0,width=0.48\textwidth]{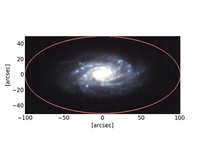}
   \includegraphics[angle=0,width=0.48\textwidth]{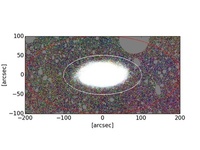}
   \includegraphics[angle=0,width=0.48\textwidth]{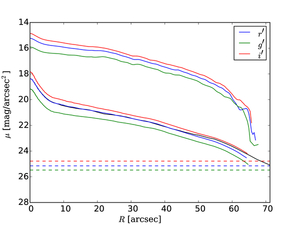}
  \caption{\label{fig:NGC4480}NGC 4480: PAS is using all quadrants.}
\end{figure}

\begin{figure}
\centering
   \includegraphics[angle=0,width=0.48\textwidth]{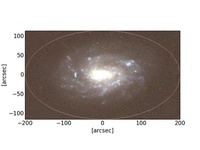}
   \includegraphics[angle=0,width=0.48\textwidth]{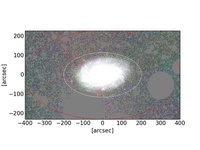}
   \includegraphics[angle=0,width=0.48\textwidth]{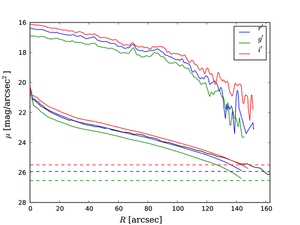}
  \caption{\label{fig:NGC4517A}NGC 4517A: PAS is using all quadrants.}
\end{figure}

\begin{figure}
\centering
   \includegraphics[angle=0,width=0.48\textwidth]{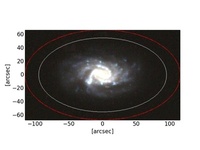}
   \includegraphics[angle=0,width=0.48\textwidth]{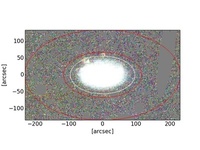}
   \includegraphics[angle=0,width=0.48\textwidth]{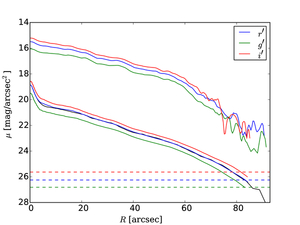}
  \caption{\label{fig:NGC4545}NGC 4545: PAS is using all quadrants.}
\end{figure}

\begin{figure}
\centering
   \includegraphics[angle=0,width=0.48\textwidth]{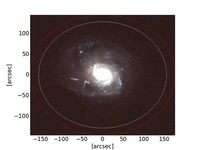}
   \includegraphics[angle=0,width=0.48\textwidth]{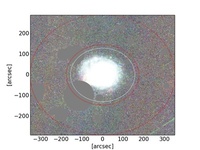}
   \includegraphics[angle=0,width=0.48\textwidth]{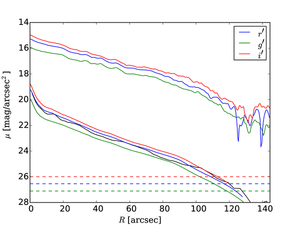}
  \caption{\label{fig:NGC4653}NGC 4653: PAS is using the PAS is using the lower right and upper right quadrants.}
\end{figure}

\begin{figure}
\centering
   \includegraphics[angle=0,width=0.48\textwidth]{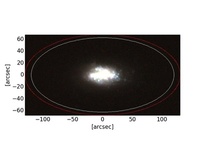}
   \includegraphics[angle=0,width=0.48\textwidth]{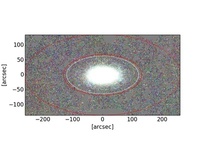}
   \includegraphics[angle=0,width=0.48\textwidth]{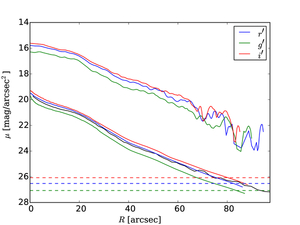}
  \caption{NGC 4668: PAS is using all quadrants}
\end{figure}

\begin{figure}
\centering
   \includegraphics[angle=0,width=0.48\textwidth]{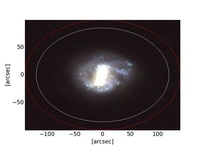}
   \includegraphics[angle=0,width=0.48\textwidth]{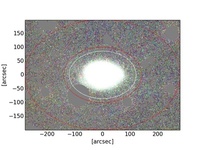}
   \includegraphics[angle=0,width=0.48\textwidth]{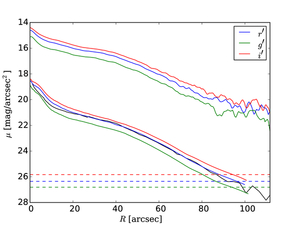}
  \caption{NGC 4904: PAS is using all quadrants.}
\end{figure}

\begin{figure}
\centering
   \includegraphics[angle=0,width=0.48\textwidth]{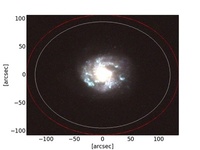}
   \includegraphics[angle=0,width=0.48\textwidth]{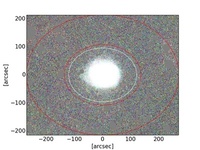}
   \includegraphics[angle=0,width=0.48\textwidth]{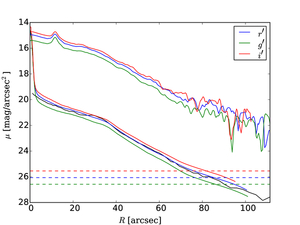}
  \caption{NGC 5147: PAS is using all quadrants.}
\end{figure}

\begin{figure}
\centering
   \includegraphics[angle=0,width=0.48\textwidth]{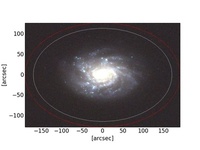}
   \includegraphics[angle=0,width=0.48\textwidth]{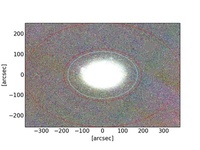}
   \includegraphics[angle=0,width=0.48\textwidth]{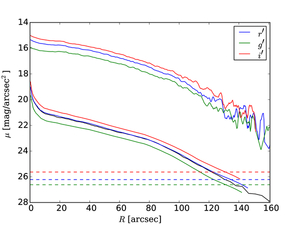}
  \caption{\label{fig:NGC5300}NGC 5300: PAS is using all quadrants.}
\end{figure}

\begin{figure}
\centering
   \includegraphics[angle=0,width=0.48\textwidth]{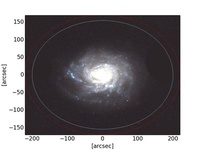}
   \includegraphics[angle=0,width=0.48\textwidth]{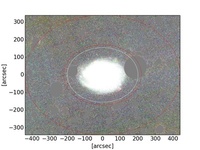}
   \includegraphics[angle=0,width=0.48\textwidth]{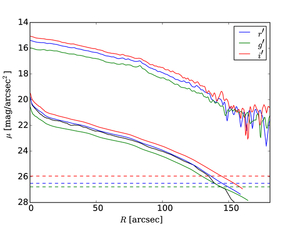}
  \caption{NGC 5334: PAS is using the PAS is using the upper left and upper right quadrants.}
\end{figure}
 
\clearpage

\begin{figure}
\centering
   \includegraphics[angle=0,width=0.48\textwidth]{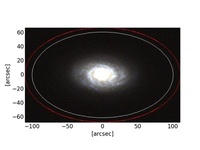}
   \includegraphics[angle=0,width=0.48\textwidth]{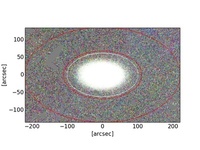}
   \includegraphics[angle=0,width=0.48\textwidth]{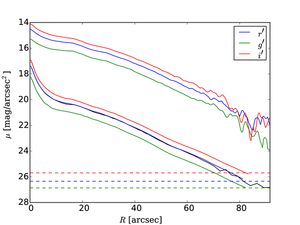}
  \caption{NGC 5376: PAS is using all quadrants.}
\end{figure}

\begin{figure}
\centering
   \includegraphics[angle=0,width=0.48\textwidth]{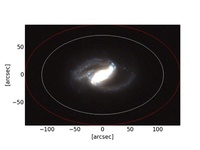}
   \includegraphics[angle=0,width=0.48\textwidth]{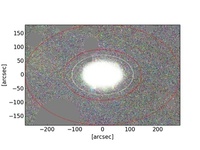}
   \includegraphics[angle=0,width=0.48\textwidth]{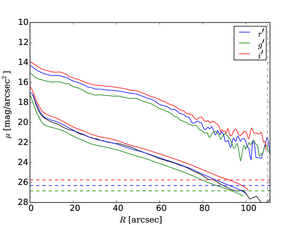}
  \caption{\label{fig:NGC5430}NGC 5430: PAS is using all quadrants.}
\end{figure}

\begin{figure}
\centering
   \includegraphics[angle=0,width=0.48\textwidth]{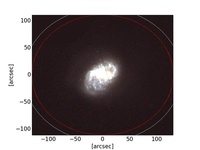}
   \includegraphics[angle=0,width=0.48\textwidth]{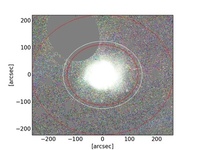}
   \includegraphics[angle=0,width=0.48\textwidth]{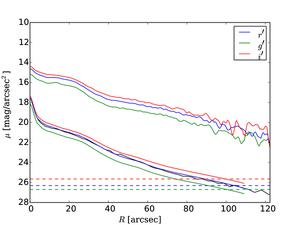}
  \caption{NGC 5480: PAS is using all quadrants.}
\end{figure}

\begin{figure}
\centering
   \includegraphics[angle=0,width=0.48\textwidth]{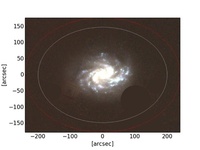}
   \includegraphics[angle=0,width=0.48\textwidth]{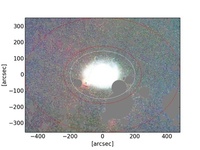}
   \includegraphics[angle=0,width=0.48\textwidth]{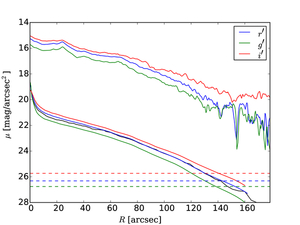}
  \caption{NGC 5584: PAS is using all quadrants.}
\end{figure}

\begin{figure}
\centering
   \includegraphics[angle=0,width=0.48\textwidth]{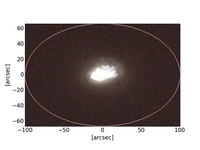}
   \includegraphics[angle=0,width=0.48\textwidth]{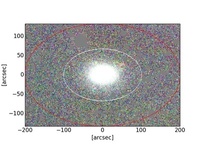}
   \includegraphics[angle=0,width=0.48\textwidth]{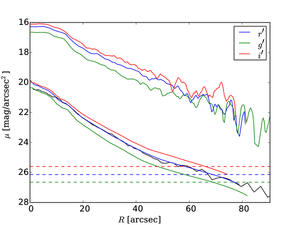}
  \caption{NGC 5624: PAS is using all quadrants.}
\end{figure}

\begin{figure}
\centering
   \includegraphics[angle=0,width=0.48\textwidth]{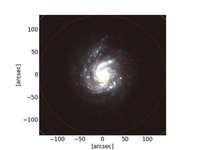}
   \includegraphics[angle=0,width=0.48\textwidth]{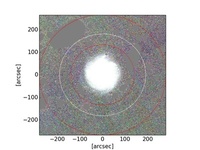}
   \includegraphics[angle=0,width=0.48\textwidth]{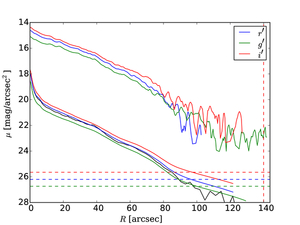}
  \caption{NGC 5660: PAS is using all quadrants.}
\end{figure}
 
\clearpage

\begin{figure}
\centering
   \includegraphics[angle=0,width=0.48\textwidth]{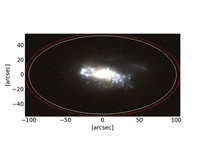}
   \includegraphics[angle=0,width=0.48\textwidth]{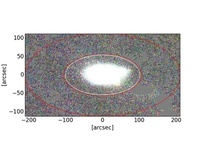}
   \includegraphics[angle=0,width=0.48\textwidth]{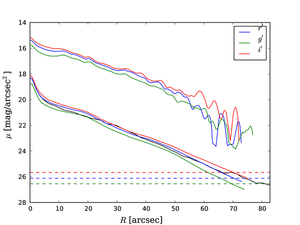}
  \caption{NGC 5667: PAS is using the upper left, lower right and upper right quadrants.}
\end{figure}

\begin{figure}
\centering
   \includegraphics[angle=0,width=0.48\textwidth]{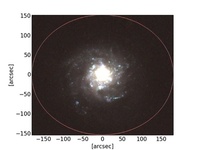}
   \includegraphics[angle=0,width=0.48\textwidth]{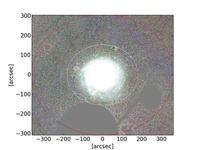}
   \includegraphics[angle=0,width=0.48\textwidth]{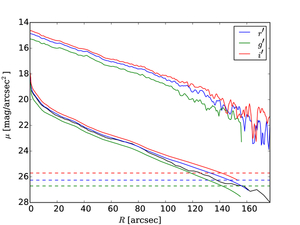}
  \caption{NGC 5668: PAS is using all quadrants.}
\end{figure}

\begin{figure}
\centering
   \includegraphics[angle=0,width=0.48\textwidth]{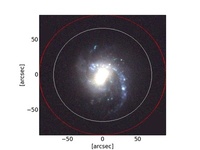}
   \includegraphics[angle=0,width=0.48\textwidth]{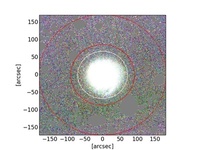}
   \includegraphics[angle=0,width=0.48\textwidth]{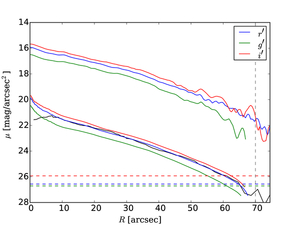}
  \caption{\label{fig:NGC5693}NGC 5693: PAS is using all quadrants.}
\end{figure}

\begin{figure}
\centering
   \includegraphics[angle=0,width=0.48\textwidth]{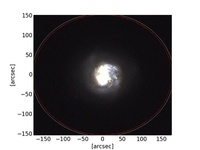}
   \includegraphics[angle=0,width=0.48\textwidth]{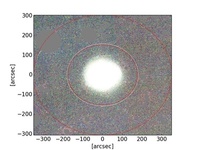}
   \includegraphics[angle=0,width=0.48\textwidth]{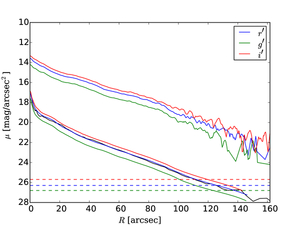}
  \caption{NGC 5713: PAS is using all quadrants.}
\end{figure}

\begin{figure}
\centering
   \includegraphics[angle=0,width=0.48\textwidth]{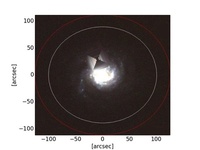}
   \includegraphics[angle=0,width=0.48\textwidth]{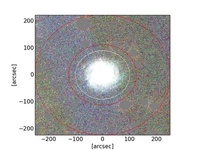}
   \includegraphics[angle=0,width=0.48\textwidth]{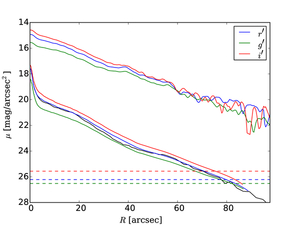}
  \caption{NGC 5768: PAS is using all quadrants.}
\end{figure}

\begin{figure}
\centering
   \includegraphics[angle=0,width=0.48\textwidth]{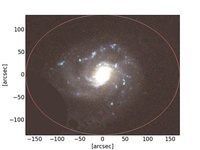}
   \includegraphics[angle=0,width=0.48\textwidth]{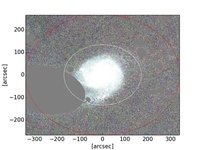}
   \includegraphics[angle=0,width=0.48\textwidth]{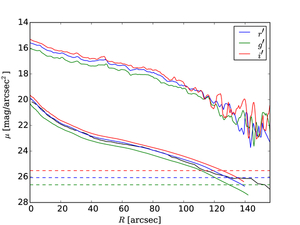}
  \caption{NGC 5774: PAS is using the lower left, lower right and upper right quadrants.}
\end{figure}

\begin{figure}
\centering
   \includegraphics[angle=0,width=0.48\textwidth]{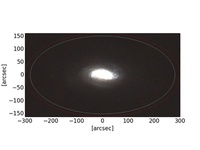}
   \includegraphics[angle=0,width=0.48\textwidth]{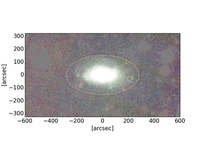}
   \includegraphics[angle=0,width=0.48\textwidth]{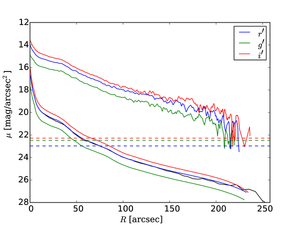}
  \caption{NGC 5806: PAS is using all quadrants.}
\end{figure}

\begin{figure}
\centering
   \includegraphics[angle=0,width=0.48\textwidth]{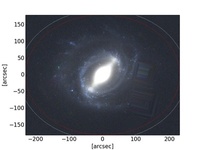}
   \includegraphics[angle=0,width=0.48\textwidth]{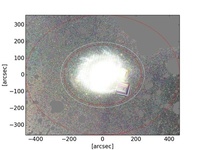}
   \includegraphics[angle=0,width=0.48\textwidth]{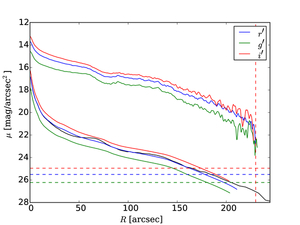}
  \caption{NGC 5850: PAS is using all quadrants.}
\end{figure}

\begin{figure}
\centering
   \includegraphics[angle=0,width=0.48\textwidth]{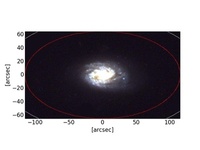}
   \includegraphics[angle=0,width=0.48\textwidth]{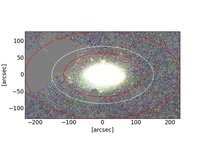}
   \includegraphics[angle=0,width=0.48\textwidth]{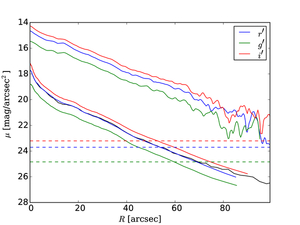}
  \caption{NGC 5937: PAS is not using the top left quadrant.}
\end{figure}

\begin{figure}
\centering
   \includegraphics[angle=0,width=0.48\textwidth]{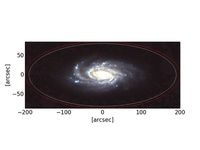}
   \includegraphics[angle=0,width=0.48\textwidth]{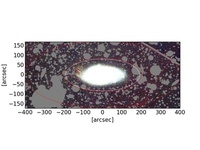}
   \includegraphics[angle=0,width=0.48\textwidth]{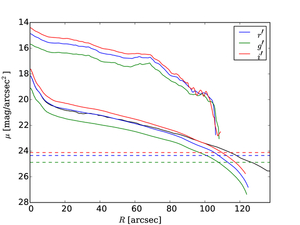}
  \caption{NGC 6070: PAS is  using all quadrants. }
\end{figure}

\clearpage

\begin{figure}
\centering
   \includegraphics[angle=0,width=0.48\textwidth]{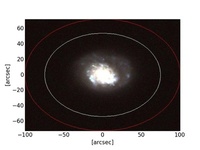}
   \includegraphics[angle=0,width=0.48\textwidth]{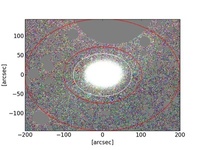}
   \includegraphics[angle=0,width=0.48\textwidth]{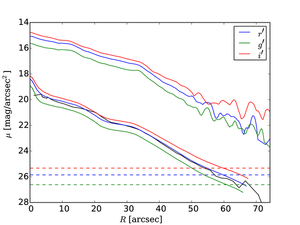}
  \caption{NGC 6155: PAS is using all quadrants.}
\end{figure}

\begin{figure}
\centering
   \includegraphics[angle=0,width=0.48\textwidth]{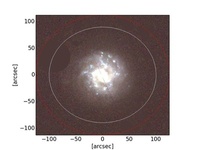}
   \includegraphics[angle=0,width=0.48\textwidth]{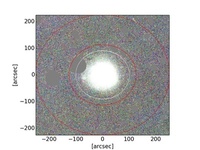}
   \includegraphics[angle=0,width=0.48\textwidth]{Petersetal-VIfigA36c.jpg}
  \caption{NGC 7437: PAS is using the upper left, lower right and upper right quadrants.}
\end{figure}

\begin{figure}
\centering
   \includegraphics[angle=0,width=0.48\textwidth]{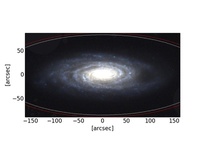}
   \includegraphics[angle=0,width=0.48\textwidth]{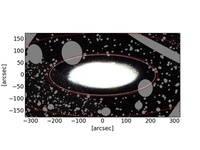}
   \includegraphics[angle=0,width=0.48\textwidth]{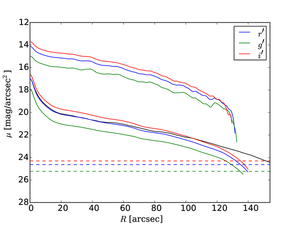}
  \caption{\label{fig:BGC7606}NGC 7606: PAS is using all quadrants.}
\end{figure}

\begin{figure}
\centering
   \includegraphics[angle=0,width=0.48\textwidth]{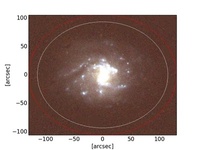}
   \includegraphics[angle=0,width=0.48\textwidth]{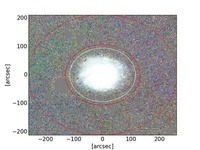}
   \includegraphics[angle=0,width=0.48\textwidth]{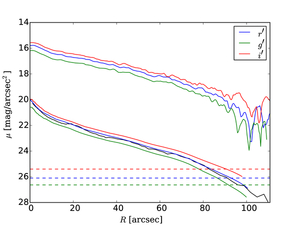}
  \caption{\label{fig:PGC006667}PGC 006667: PAS is using all quadrants.}
\end{figure}

\begin{figure}
\centering
   \includegraphics[angle=0,width=0.48\textwidth]{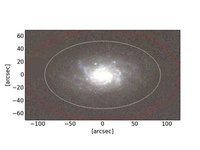}
   \includegraphics[angle=0,width=0.48\textwidth]{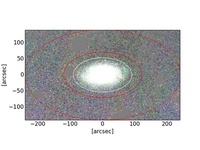}
   \includegraphics[angle=0,width=0.48\textwidth]{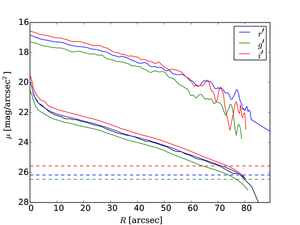}
  \caption{UGC 02081: PAS is using all quadrants.}
\end{figure}

\begin{figure}
\centering
   \includegraphics[angle=0,width=0.48\textwidth]{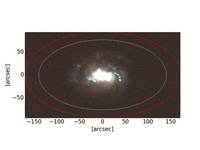}
   \includegraphics[angle=0,width=0.48\textwidth]{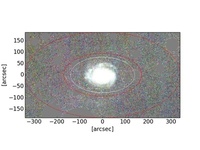}
   \includegraphics[angle=0,width=0.48\textwidth]{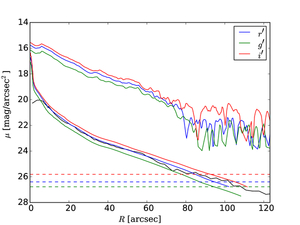}
  \caption{UGC 04393: PAS is using all quadrants.}
\end{figure}

\begin{figure}
\centering
   \includegraphics[angle=0,width=0.48\textwidth]{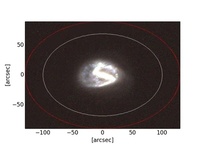}
   \includegraphics[angle=0,width=0.48\textwidth]{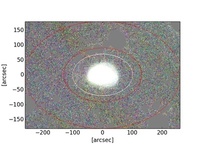}
   \includegraphics[angle=0,width=0.48\textwidth]{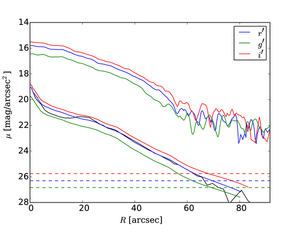}
  \caption{UGC 06309: PAS is using all quadrants.}
\end{figure}

\begin{figure}
\centering
   \includegraphics[angle=0,width=0.48\textwidth]{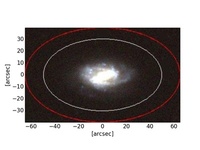}
   \includegraphics[angle=0,width=0.48\textwidth]{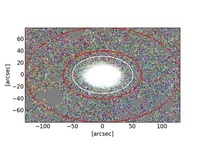}
   \includegraphics[angle=0,width=0.48\textwidth]{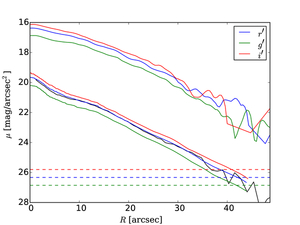}
  \caption{UGC 06518: PAS is using all quadrants.}
\end{figure}

\begin{figure}
\centering
   \includegraphics[angle=0,width=0.48\textwidth]{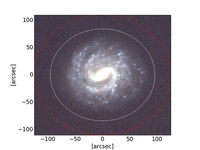}
   \includegraphics[angle=0,width=0.48\textwidth]{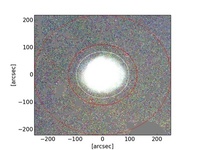}
   \includegraphics[angle=0,width=0.48\textwidth]{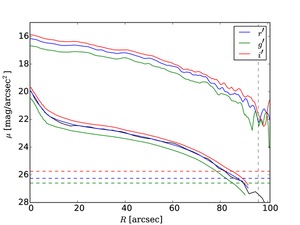}
  \caption{UGC 06903: PAS is using all quadrants.}
\end{figure}

\begin{figure}
\centering
   \includegraphics[angle=0,width=0.48\textwidth]{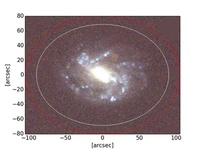}
   \includegraphics[angle=0,width=0.48\textwidth]{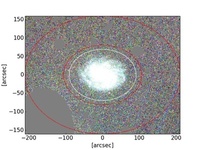}
   \includegraphics[angle=0,width=0.48\textwidth]{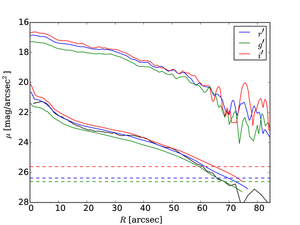}
  \caption{UGC 07700: PAS is using all quadrants.}
\end{figure}

\begin{figure}
\centering
   \includegraphics[angle=0,width=0.48\textwidth]{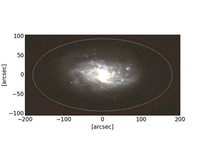}
   \includegraphics[angle=0,width=0.48\textwidth]{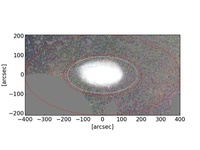}
   \includegraphics[angle=0,width=0.48\textwidth]{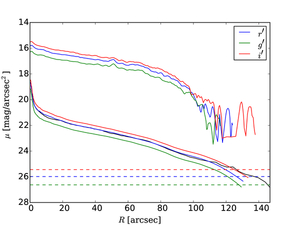}
  \caption{\label{fig:UGC08041}UGC 08041: PAS is not using lower left quadrant.}
\end{figure}

\begin{figure}
\centering
   \includegraphics[angle=0,width=0.48\textwidth]{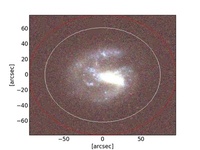}
   \includegraphics[angle=0,width=0.48\textwidth]{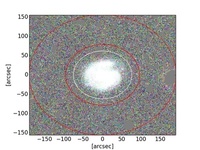}
   \includegraphics[angle=0,width=0.48\textwidth]{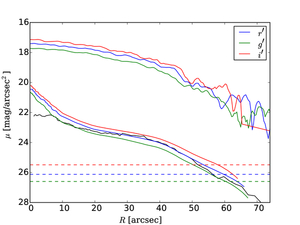}
  \caption{\label{fig:UGC08084}UGC 08084: PAS is using all quadrants.}
\end{figure}

\begin{figure}
\centering
   \includegraphics[angle=0,width=0.48\textwidth]{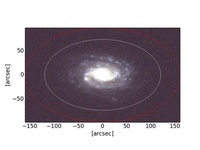}
   \includegraphics[angle=0,width=0.48\textwidth]{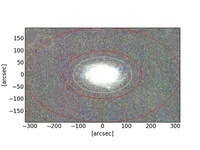}
   \includegraphics[angle=0,width=0.48\textwidth]{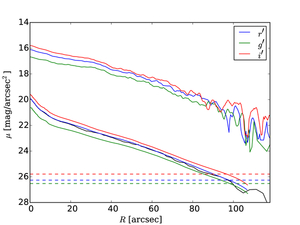}
  \caption{UGC 08658: PAS is using all quadrants.}
\end{figure}

\begin{figure}
\centering
   \includegraphics[angle=0,width=0.48\textwidth]{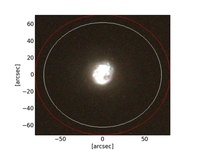}
   \includegraphics[angle=0,width=0.48\textwidth]{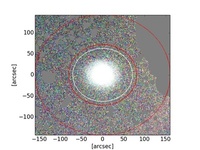}
   \includegraphics[angle=0,width=0.48\textwidth]{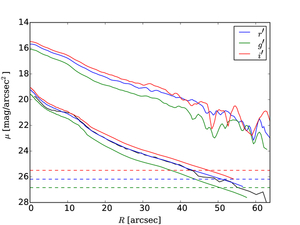}
  \caption{UGC 09741: PAS is using all quadrants.}
\end{figure}

\begin{figure}
\centering
   \includegraphics[angle=0,width=0.48\textwidth]{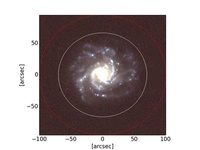}
   \includegraphics[angle=0,width=0.48\textwidth]{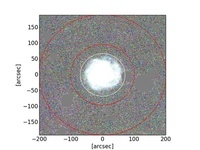}
   \includegraphics[angle=0,width=0.48\textwidth]{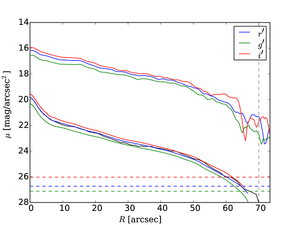}
  \caption{UGC 09837: PAS is using all quadrants.}
\end{figure}

\begin{figure}
\centering
   \includegraphics[angle=0,width=0.48\textwidth]{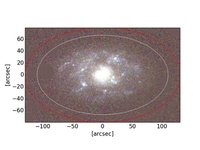}
   \includegraphics[angle=0,width=0.48\textwidth]{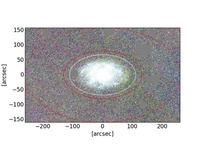}
   \includegraphics[angle=0,width=0.48\textwidth]{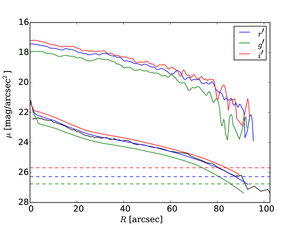}
  \caption{\label{fig:UGC12709}UGC 12709: PAS is using all quadrants.}
\end{figure}

\twocolumn

\bsp

\label{lastpage}

\end{document}